\documentclass[aoas,MSNbibl,nameyear,seceqn,rotating,dvips]{arximspdf}
\usepackage{dcolumn}
\usepackage{stfloats}
\usepackage{graphicx}

\doi{10.1214/13-AOAS695} 
\volume{8}
\issue{1}
\pubyear{2014}
\firstpage{176}
\lastpage{203}

\makeatletter
  \fnbelowfloat
\newcolumntype{d}[1]{D{.}{.}{#1}}
\newtheorem{prop}{Proposition}[section]
\newtheorem{note}{Note}[section]
\makeatother

\begin{document}
\begin{frontmatter}

\title{Bayesian methods for genetic association analysis with
heterogeneous subgroups:
From meta-analyses to gene--environment interactions\thanksref{T1}}
\runtitle{Genetic association analysis with heterogeneous subgroups}

\begin{aug}
\author[A]{\fnms{Xiaoquan} \snm{Wen}\corref{}\ead[label=e1]{xwen@umich.edu}}
\and
\author[B]{\fnms{Matthew} \snm{Stephens}\ead[label=e2]{mstephens@uchicago.edu}}
\runauthor{X. Wen and M. Stephens}
\affiliation{University of Michigan and University of Chicago}
\address[A]{Department of Biostatistics\\
University of Michigan\\
1415 Washington Heights\\
Ann Arbor, Michigan 48109\\
USA\\
\printead{e1}} 
\address[B]{Department of Statistics\\
and\\
Department of Human Genetics\\
University of Chicago\\
5734 S. University Avenue\\
Chicago, Illinois 60637\\
USA\\
\printead{e2}}
\end{aug}
\thankstext{T1}{Supported in part by NIH Grants HG02585 to M. Stephens
and MH090951-02 (PI Jonathan Pritchard).}

\received{\smonth{11} \syear{2011}}
\revised{\smonth{10} \syear{2013}}

%
\begin{abstract}
Genetic association analyses often involve data from multiple
potentially-heterogeneous subgroups.
The expected amount of heterogeneity can vary from modest (e.g., a
typical meta-analysis)
to large (e.g., a strong gene--environment interaction). However,
existing statistical tools
are limited in their ability to address such heterogeneity. Indeed,
most genetic association
meta-analyses use a ``fixed effects'' analysis, which assumes no
heterogeneity. Here we develop and apply Bayesian association methods
to address this problem.
These methods are easy to apply (in the simplest case, requiring only a
point estimate
for the genetic effect and its standard error, from each subgroup) and
effectively
include standard frequentist meta-analysis methods, including the usual
``fixed effects'' analysis, as special cases.
We apply these tools to two large genetic association studies: one a
meta-analysis of
genome-wide association studies from the Global Lipids consortium, and
the second a
cross-population analysis for expression quantitative trait loci
(eQTLs). In the
Global Lipids data we find, perhaps surprisingly, that effects are
generally quite
homogeneous across studies. In the eQTL study we find that eQTLs are
generally shared among different
continental groups, and discuss consequences of this for study design.
\end{abstract}

%

\end{frontmatter}

\setcounter{footnote}{1}

\section{Introduction}\label{sec1}
We consider the following problem, which arises
frequently in genetic association analysis: how to test for association
while allowing for heterogeneity
of effects among subgroups. We are motivated particularly by the
following applications:

\subsection*{Motivating application 1: The Global Lipids genome-wide association study}
The Global Lipids consortium [\citet{Teslovich2010}] conducted a large
meta-analysis of genome-wide genetic association studies of blood
lipids phenotypes [total cholesterol (TC), low-density lipoprotein
cholesterol (LDL-C), high-density lipoprotein cholesterol (HDL-C) and
triglycerides (TG)].
This study, like most meta-analyses, aimed to increase power by
combining information across studies. The consortium amassed a total of
more than 100,000 individuals, through 46~separate studies. These
studies involve different investigators, at different centers, with\vadjust{\goodbreak}
different enrollment criteria. Consequently one would expect genetic
effect sizes to differ among studies. However, \citet{Teslovich2010},
following standard practice in this field, analyzed the data assuming
no heterogeneity. This analysis appeared highly successful, identifying
genetic associations at a total of~95 different genetic loci, 53 of
them novel. Our work here was motivated by
a desire to analyze these data, and others like them,
\textit{taking account of potential heterogeneity among studies}, and
to see whether this would identify additional
genetic associations.

\subsection*{Motivating application 2: Assessing heterogeneity of genetic effects
on gene expression (eQTLs) among populations}
An expression quantitative trait locus (eQTL) is a genetic variant that
is associated with expression (activity) of a gene. Identifying eQTLs is
important because such variants are candidates for being functional
(i.e., actually causing changes in gene expression), and hence
candidates for having other, perhaps medically important, consequences.
See \citet{yoavjonathan} for more on insights to be gained from eQTL studies.

Here we analyze data from \citet{Stranger2007}, who measured gene
expression on lymphoblastoid cell lines derived from unrelated
individuals sampled from three major continental groups: Europeans
(CEU), Asians (ASN), and Africans (YRI). The main aim of our analysis
is to
assess heterogeneity of eQTL effects across groups: for example, do
eQTL effects tend to vary among groups, and do some eQTLs appear to be
active in only some groups? Understanding heterogeneity in this context
could yield insights into differences in the gene-regulatory \mbox{mechanisms}
acting in each group, and also has important implications for generalizability
of studies performed in one subgroup to other subgroups.
Similar questions also arise frequently in eQTL studies involving
different tissue or cell types [\citet{Dimas2009a}, \citet{casey},
\citet{flutre}].


\subsection*{Motivating application 3: Identifying biological
interactions with environment}

Our third example is more generic, but nonetheless important:
identifying genetic associations
in the presence of environmental interactions. Strong environmental
interactions can result in genetic effects varying among subgroups, and
in extreme cases could even cause effects to have different signs in
different subgroups. In such cases ignoring heterogeneity would
substantially reduce power. For example, by separately analyzing male
and females, \citet{Kong2008} identified a strong genetic association
with recombination rate that is missed by a standard analysis that
ignores heterogeneity. However, separately analyzing subgroups is less
attractive than analyzing them jointly and allowing for potential
heterogeneity; this motivated development of methods described here.\vspace*{9pt}

These three applications differ in their expected heterogeneity. For
example, whereas interactions could cause genetic effects to differ in
sign among subgroups, differences in sign
seem less likely in the meta-analysis setting. However, they also share
an important element in common: the vast majority of genetic variants\vadjust{\goodbreak}
are unassociated with any given phenotype within \textit{all} subgroups.
Consequently, it is of considerable interest to identify genetic
variants that show association in \textit{any} subgroup or, in other words
to reject the ``global'' null hypothesis of \textit{no association within
any subgroup}. This focus on rejecting the global null hypothesis
distinguishes genetic association analyses from other settings and
calls for analysis approaches tailored to this goal; see \citet{Lebrec2010} for relevant discussion. Thus, although there
has been previous work on Bayesian methods for meta-analysis [e.g.,
\citet{Sutton2001}, \citet{stangelberrybook}, \citet{DuMouchel1983},
\citet{Whitehead1991}, \citet{Li1994}, \citet{Eddy1990},
\citet{Givens1997}, \citet{Verzilli2008}, \citet{DeIorio2011},
\citet{Burgess2010}, \citet{Mila2011}],
the nature of our applications calls for a different focus.
Specifically, our applications need tools for association testing and
model comparison, via Bayes Factors (BFs), rather than for estimation.
In addition, because genetic association studies often involve very
many association tests, computational speed is important, so we focus
on obtaining fast numerical approximations to BFs (rather than using
MCMC say). Finally, because in many cases, including the Global Lipids
data above, only summary data (e.g., effect size estimates and standard
errors in the different subgroups) are easily available, we need
methods that can work with summary data.

Although our methods are Bayesian, they have
close connections with related frequentist methods. Indeed,
while this work was in progress, \citet{Lebrec2010} [and, later, \citet{Han2011}] published
frequentist approaches to association testing based on
models very similar to those used here.
Further, as we show in Section~\ref{secanalytic}, some standard
frequentist meta-analysis tests correspond closely to BFs
obtained under certain priors. Consequently, ranking SNP associations
by these standard test statistics is equivalent
to making specific (and not necessarily realistic) prior assumptions.
Thus, although our primary goal is to provide a practical solution to a common
applied problem, we also provide theoretical results linking our methods
to widely used existing methods.

\section{Models and methods}\label{sec2}

The problems outlined above have two key goals:
\begin{enumerate}
\item To test whether a genetic variant is associated with phenotype in
any subgroup, allowing for potential heterogeneity of effects among subgroups.
\item Given that an association exists, assess the support for
different levels of heterogeneity.
\end{enumerate}
We tackle these problems by specifying a family of alternative models
with varying levels of heterogeneity, and by developing computational
tools to calculate the support in the data (the BF) for
each alternative model compared with the null model of no association.
Within this framework the goal of testing the global null (1 above) is
accomplished by assessing the overall support for any of the
alternative hypotheses, whereas the goal of examining heterogeneity among
groups is achieved by comparing the relative support for different
alternative models.\vadjust{\goodbreak}

All our motivating examples involve quantitative outcomes, and we focus
on this case. However, in common to many meta-analysis methods, the
simplest of our methods requires only an estimated effect size in each
study and its corresponding standard deviation, and thus can
be applied to any setting where such estimates are available
(e.g., generalized linear models). See Appendix~B of the supplementary material [\citet{suppA}] for details.

\subsection{Notation and assumptions}\label{sec2.1}

Assume quantitative phenotype data and genotype data are available on
$S$ predefined subgroups. Like most association analyses, we analyze
each genetic variant in turn, one at a time.
Assume that the data within subgroup $s$ come from $n_s$ randomly
sampled unrelated individuals. Let the $n_s$-vectors $\mathbf{y}_s$ and $\mathbf{g}_s$ denote, respectively, the
corresponding phenotype and genotype data, and $\mathbf{Y}:= (\mathbf{y}_1,\ldots,\mathbf{y}_S)$ and $\mathbf{G}:=
(\mathbf{g}_1,\ldots,\mathbf{g}_S)$. Here each genotype
is coded as 0, 1 or 2 copies of a reference allele, so $\mathbf{g}_{s} \in\{0,1,2\}^{n_s}$. (For imputed genotypes we
replace each genotype with its posterior mean [\citet{guanstephens}],
in which case $\mathbf{g}_{s} \in[0,2]^{n_s}$.)

\subsection{Models for effect-size heterogeneity}\label{sec2.2}

Within each subgroup, we model genotype--phenotype association using a
standard linear model:
%
\begin{equation}
\label{studylinmodel} \mathbf{y}_s = \mu_s{\mathbf1} +
\beta_s \mathbf{g}_s + \mathbf{e}_s,\qquad
\mathbf{e}_s \sim\mathrm{N}\bigl(0, \sigma_s^2I
\bigr)
\end{equation}
with residual errors $\mathbf{e}_s$ assumed independent across subgroups.
[Additional, possibly study-specific, covariates are easily added to
the right-hand side of (\ref{studylinmodel}).
If independent flat priors are used for the coefficients of these
covariates within each study, then our main results below still hold,
effectively unchanged. This treatment is analogous to the frequentist
mixed-effects model, where such covariates are typically assumed to
have study-specific effects.]

The ``global'' null hypothesis is no genotype--phenotype association
within any subgroup, that is, $\beta_s=0$ for all $s$.

Under the alternative hypothesis the genetic effects are nonzero. To
allow for heterogeneity among subgroups, we assume that these effects
are normally distributed about some unknown common mean. We consider
two different \mbox{definitions} of genetic effects: the ``standardized
effects'', $b_s:= \beta_s/\sigma_s$, and the unstandardized effects,
$\beta_s$, leading to the following models:

\begin{longlist}[2.]
\item[1.] \textit{Exchangeable standardized effects} (\textit{ES model}). The
standardized effects $b_s$ are normally distributed among subgroups,
about some unknown mean $\bar b$, to which we assign a normal prior:
%
\begin{eqnarray}\label{priorhbs}
b_s | \bar b, \sigma_s & \sim&\mathrm{N} \bigl( \bar b, \phi^2\bigr)\quad\bigl[\mbox{or, equivalently, }
\beta_s | \sigma_s \sim\mathrm{N}\bigl(
\sigma_s \bar b, \sigma_s^2
\phi^2\bigr)\bigr],
\\
\label{priorhbbar} \bar b &\sim&\mathrm{N}\bigl(0, \omega^2\bigr).
\end{eqnarray}
Alternatively, and equivalently, the vector $\mathbf{b}=(b_1,\ldots,b_S)$ is
multivariate normally distributed:
%
\begin{equation}
\mathbf{b}\sim\mathrm{N}_S(0, \Sigma_b),
\end{equation}
where $\Sigma_b$ is an $S \times S$ matrix with diagonal elements
$\operatorname{Var}(b_s)=\phi^2+\omega^2$ and off-diagonal elements
$\operatorname{Cov}(b_s,
b_{s'}) = \omega^2$.

\item[2.] \textit{Exchangeable effects} (\textit{EE model}). The unstandardized
effects $\beta_s$ are normally distributed about some unknown mean
$\bar\beta$, to which we assign a normal prior:
%
\begin{eqnarray}
\label{priorsbs} \beta_s | \bar\beta& \sim&\mathrm{N}\bigl( \bar\beta,
\psi^2\bigr),
\\
\bar\beta& \sim&\mathrm{N}\bigl(0, w^2\bigr).
\end{eqnarray}
Alternatively, and equivalently, the vector $\bolds{\beta}=(\beta
_1,\ldots,\beta_S)$ is multivariate normally distributed:
%
\begin{equation}
\bolds{\beta}\sim\mathrm{N}_S(0, \Sigma_\beta),
\end{equation}
where $\Sigma_\beta$ is an $S \times S$ matrix with diagonal elements
$\operatorname{Var}(\beta_s)=\psi^2+w^2$ and off-diagonal elements
$\operatorname{Cov}(\beta
_s, \beta_{s'}) = w^2$.
\end{longlist}

In both ES and EE models we assume conjugate priors for $(\mu_s,
\sigma_s)$:
%
\begin{eqnarray}
\mathrm{ES}\dvtx \label{priorhmu} \mu_s | \sigma_s &\sim&
\mathrm{N}\bigl(0, \sigma_s^2 u_s^2
\bigr);\qquad  \sigma _s^{-2} \sim\Gamma(m_s/2,
l_s/2),
\\
\mathrm{EE}\dvtx   \mu_s &\sim&\mathrm{N}\bigl(0, u_s^2
\bigr);\qquad \sigma_s^{-2} \sim\Gamma( m_s/2,
l_s/2).
\end{eqnarray}
Specifically, we consider posteriors and BFs that arise in the limits
$u_s^2 \rightarrow\infty$ and $l_s, m_s \rightarrow0$.
These limiting priors correspond to standard improper priors for normal
regressions and ensure that the BFs satisfy certain invariance properties
[see \citet{servinstephens}].

Both ES and EE have two key hyperparameters, one ($\omega$ in ES; $w$
in EE) that controls the prior expected size of the \textit{average
effect} across subgroups, and another ($\phi$ in ES; $\psi$ in EE)
that controls the prior expected degree of \textit{heterogeneity} among
subgroups. A complimentary view is that $\omega^2 + \phi^2$
(resp., $w^2 + \psi^2$) controls the expected (marginal) effect
size in each study and $\phi/\omega$ (resp., $\psi/w$)
controls the degree of heterogeneity. Thus, one can allow for different
levels of heterogeneity by considering different values of these
hyperparameters (see below). Note that $\phi=0$ (resp., $\psi
=0$) corresponds to the assumption, commonly used in practice, of no
heterogeneity among subgroups.

Of the two models, ES has the advantage that it results in analyses
(e.g., BFs) that are invariant to the phenotype measurement scale used
within each subgroup. This makes it robust to users accidentally
specifying phenotype measurements in different subgroups on different
scales (a nontrivial issue in complex analyses involving collaboration
among many research groups). It also makes it applicable when
measurement scales are difficult to harmonize across subgroups, for
example, due to use of different measurement technologies. For these
reasons we prefer ES for general use. However, in some cases EE may be
easier to apply. For example, if one has access only to published point
estimates and standard errors for the effect size $\beta_s$ in each
study, then this suffices to approximate the BF under EE, but not under
ES. Note that ES and EE will produce similar results if the residual
error variances are similar in all subgroups.

\subsubsection{Limiting heterogeneity: A curved exponential family normal prior}\label{sec2.2.1}

The above priors assume independence of the mean ($\bar b$) and
variance ($\omega^2$) of the effects.
In some settings this assumption may be unattractive. For example,
in a typical meta-analysis we expect effects to show ``limited
heterogeneity'' across studies and typically to have the same sign
[\citet{Owen2009}], regardless of whether $\bar b$ is small or large.
But the independence assumption implies that the probability that the
effects have the same sign is much larger when $\bar b$ is large than
when it is small. To address this, we can replace the priors (\ref
{priorhbs}) and (\ref{priorsbs}) with, respectively,
%
\begin{eqnarray}
b_s &\sim&\mathrm{N}\bigl(\bar b, k^2 {\bar
b}^2\bigr),
\\
\beta_s &\sim&\mathrm{N}\bigl(\bar\beta, k^2{\bar
\beta}{}^2\bigr).
\end{eqnarray}
Here $k$ determines the amount of heterogeneity, with smaller $k$
indicating less heterogeneity
and $k=0$ indicating no heterogeneity. Under these priors the
probability of effects differing in sign depends only on $k$ and not on
$\bar b$,
%
\begin{equation}
\label{eqnprobsamesign}
\Pr ( b_s \mbox{ has a different sign from } \bar b )
= \Phi\biggl( -\frac{1}{|k|}\biggr),
\end{equation}
where $\Phi$ is the cumulative distribution function of a standard
normal distribution.
For example, when $k=1/2$, (\ref{eqnprobsamesign}) is approximately
2.3\%.

We call these priors ``Curved Exponential Family Normal'' (CEFN) priors,
reflecting their functional relationship between the mean and variance.

\subsection{Bayes factors for testing the global null hypothesis}\label{sec2.3}\label{bfcompsection}

For simplicity we focus on calculations for the ES model;
details for EE, and modifications for CEFN, are given in the appendices [supplementary material \citet{suppA}].

The ES model has two hyperparameters, $\phi$ and $\omega$.
The global null hypothesis, which is most naturally written as $\beta
_s \equiv0$ for all $s$, can also be written as
%
\begin{equation}
H_0\dvtx  \phi= \omega= 0.
\end{equation}
The support in the data for a particular alternative model, specified
by hyperparameters $(\phi, \omega)$, vs $H_0$, is given
by the Bayes Factor (BF):
%
\begin{equation}
\mathrm{BF}^{\mathrm{ES}}(\phi, \omega) = \frac{P(\mathbf{Y}|
\mathbf{G}, \phi,\omega) }{
P(\mathbf{Y}|\mathbf{G},H_0)}.
\end{equation}

Each value of $(\omega,\phi)$ corresponds to a particular alternative
model, with $\omega$ controlling the typical average effect size, and
$\phi$ controlling the degree of heterogeneity among subgroups (or in
a reparametrization, $\omega^2 + \phi^2$ controls the expected
marginal effect size in each subgroup and $\phi/\omega$ controls the
degree of heterogeneity). To allow for uncertainty about appropriate
values for $\phi$ and $\omega$, we use a (discrete) prior
distribution on a set of plausible values $\{ (\phi^{(i)}, \omega
^{(i)})\dvtx  i=1,\ldots,M\}$ (see applications for details). (This induces
a prior on the effects that is a mixture of multivariate normals.) A
discrete prior is more flexible than fixing $\phi,\omega$ to specific
values, while maintaining computational convenience.
Indeed, if $\pi_i$ denotes the prior weight on $(\phi^{(i)}, \omega
^{(i)})$, then the resulting BF against $H_0$ is the weighted average
of the individual BFs:
%
\begin{equation}
\label{bfmeta} \mathrm{BF}^{\mathrm{ES}}_{\mathrm{av}}:= \sum
_{i=1}^M \pi_i \mathrm{BF}^{\mathrm{ES}}
\bigl(\phi^{(i)}, \omega^{(i)}\bigr).
\end{equation}
This average could be extended to include other models (e.g., some
using the CEFN prior, others not).
The fact that BFs under different assumptions for heterogeneity
can be both averaged in this way (to assess evidence against the global
null, allowing for heterogeneity), and compared with one another (to
assess the evidence for different levels or types of heterogeneity), is
one nice feature of the Bayesian framework.

We make two comments regarding the need to specify the prior weights
$\pi$ in (\ref{bfmeta}).
First, the usual practice of simply ignoring heterogeneity
is implicitly making a particular, and rather strong, decision about
$\pi$.
In this sense, specifying weights for different amounts of
heterogeneity is simply turning a usually implicit decision into an
explicit decision.
[See \citet{Wakefield2009} for analogous discussion regarding choice of
priors on effect sizes.] Second, in some applications it will be
possible, and desirable, to learn these weights from the data via a
hierarchical model,
reducing the subjectivity of the analysis. Indeed, the ability of BFs
to be naturally incorporated into a hierarchical model is one advantage
over analogous frequentist test statistics [\citet{Lebrec2010}]
that maximize over $\phi,\omega$. This idea is illustrated in
Section~\ref{popeqtl}.

\subsubsection{Calculating Bayes factors}\label{sec2.3.1}
Calculating $\mathrm{BF}^{\mathrm{ES}}(\phi, \omega)$ involves
evaluating a
multidimensional integral. In Appendix~A of the supplementary material [\citet{suppA}] we present two
different approximations, both based on applying Laplace's method and
both having error terms that decay inversely with the average sample
size across subgroups. The first of these, which effectively follows
methods from \citet{Butler2002} for computing confluent hyper-geometric
functions, is very accurate, even for small sample sizes. Indeed, for
the special case of a single subgroup ($S=1$), the approximation
becomes \textit{exact}, and for small numbers of subgroups we have checked
numerically (Appendix~D of the supplementary material [\citet{suppA}]) that it provides almost
identical results to an alternative approach based on adaptive
quadrature (which is practical only for small $S$). However, it
requires a numerical optimization step and has a somewhat complex form,
which, although not a practical barrier to its use, does hinder
intuitive interpretation. In what follows we use $\widehat{\mathrm{BF}}{}^{\mathrm{ES}}$ to denote
this approximation.

The second approximation is less accurate for small samples sizes, but
converges asymptotically (with average sample size) to the correct answer.
Further, a~simple modification, described in Appendix~C of the supplementary material [\citet{suppA}], yields much greater accuracy.
For the special case of $S=1$ it yields an analogue of the approximate
BFs from \citet{Wakefield2009} and \citet{Johnson2008}, and in what
follows we use $\mathrm{ABF}^{\mathrm{ES}}$ to denote this
approximation under the ES
model. The nice feature of $\mathrm{ABF}^{\mathrm{ES}}$ is that it
has an intuitive
analytic form, which is detailed after the applications,
in Section~\ref{secanalytic} (Proposition \ref{abfesprop}).

\subsubsection{Special cases}\label{sec2.3.2}
Our ES and EE models include, in the special cases $\phi=0$ and $\psi
=0$, the case where there is no heterogeneity of effects across subgroups.
The assumption of no heterogeneity also underlies standard ``fixed
effects'' meta-analysis methods, and so
we use $\mathrm{ABF}_{\mathrm{fix}}^{\mathrm{ES}}$ and $\mathrm
{ABF}_{\mathrm{fix}}^{\mathrm{EE}}$ to denote ABFs computed under
these assumptions. These ABFs
are the Bayesian analogues of \mbox{standard} frequentist test statistics for
``fixed effect'' analyses; see Section~\ref{fixmodelsec} for details.

At the other extreme, the cases $\omega=0$ (ES model) and $w=0$ (EE
model) maximize heterogeneity across subgroups, and
we use $\mathrm{ABF}_{\max\mathrm{H}}^{\mathrm{ES}}$, $\mathrm
{ABF}_{\max\mathrm{H}}^{\mathrm{EE}}$ to denote ABFs
computed under these assumptions.
These ABFs can be viewed as the Bayesian analogues of frequentist
methods that combine information
across subgroups, ignoring the direction of the effect in each subgroup
(e.g., Fisher's method); see Section~\ref{maxhmodelsec} for details.

\subsubsection{Bayes factors from summary statistics}\label{sec2.3.3}
It\vspace*{1pt} turns out that both the true BFs ($\mathrm{BF}^{\mathrm{ES}}$,
$\mathrm{BF}^{\mathrm{EE}}$) and the
approximations $(\widehat{\mathrm{BF}}{}^{\mathrm{ES}}$, $\widehat
{\mathrm{BF}}{}^{\mathrm{EE}}$, $\mathrm{ABF}^{\mathrm{ES}}$,
$\mathrm{ABF}^{\mathrm{EE}})$ depend on the observed data in each
subgroup only through a
set of summary statistics, a~6-tuple $(n_s, {\mathbf1}'\mathbf{y}_s, {\mathbf1}'\mathbf{g}_s, \mathbf{y}_s'\mathbf{y}_s, \mathbf{g}_s'\mathbf{g}_s, \mathbf{y}_s'\mathbf{g}_s)$.
Furthermore, for the simplest approximations (the ABFs), the summary
statistics needed from each subgroup are reduced to only $(\hat b_s,\operatorname{se}(\hat b_s) )$ for the ES model and ($\hat\beta_s,
\operatorname{se}(\hat\beta_s) )$ for the EE model. These are exactly the
quantities used
in traditional meta-analysis applications.

These properties have important practical implications. First,
they aid collaboration among groups, where sharing of raw data can be
more difficult than sharing summary data. Second, and perhaps more importantly,
it means that the methods, particularly the ABFs, are extremely flexible.
Indeed, in any setting where an effect size estimate and its standard
error can be obtained for each subgroup, these can be
plugged in to compute an ABF. This can be viewed as an additional
Laplace approximation (Appendix~B of the supplementary material [\citet{suppA}]). Thus, the methods
can easily handle study-specific covariates, and nonnormal data
(e.g., using a generalized linear model within each subgroup).

\subsubsection{Software}\label{sec2.3.4}

We implemented these methods in a software package MeSH (Meta-analysis
with Subgroup Heterogeneity), available from
\texttt{\href{http://www.github.com/xqwen/mesh}{http://}
\href{http://www.github.com/xqwen/mesh}{www.github.com/xqwen/mesh}}.

\section{Applications}\label{sec3}

\subsection{Illustrative example: deCODE recombination study}\label{sec3.1}

We first illustrate the methods on motivating application~3 where
heterogeneity of effects is known to occur.
The example involves three correlated genetic variants (SNPs) that were\vadjust{\goodbreak}
found, in a genetic association
study of recombination rate by \citet{Kong2008}, to be strongly associated
with both male and female recombination rates, but with estimated
effects in opposite directions (i.e., the allele associated with lower
recombination rate in males is associated with higher recombination
rate in females).\footnote{A subsequent study suggests that this
genetic region may actually contain more than one genetic variant
affecting recombination rates, some acting in males and others in
females, rather than a single genetic variant with antagonistic effects
in the two groups [\citet{Fledel-Alon2011}].
This finding emphasizes the fact that apparent heterogeneity may differ
from actual heterogeneity, particularly when examining genetic markers
that are not the causal variant. The power benefits of accounting for
heterogeneity in an association analysis largely depend on apparent,
rather than actual, heterogeneity.}

We applied our methods to these data, exploiting the ability to compute
ABFs from summary data. Specifically, Table~1 of \citet{Kong2008} gives
estimated effect sizes $\hat\beta_{\mathrm{male}} $ and $\hat\beta
_{\mathrm{female}}$ for three relevant SNPs, as well as corresponding
\mbox{$p$-}values,
from which we infer approximate values for the standard errors $\operatorname{se}( \hat\beta_{\mathrm{male}})$ and $\operatorname{se}( \hat\beta
_{\mathrm{male}})$.
These summaries suffice to compute ABFs under the EE model.
We treat males and females as two subgroups and consider 4 levels of
expected marginal overall effect sizes with $\sqrt{\psi^2+ w^2} = 5,
10, 20, 40$ (the phenotype scale being centi-Morgans) and 5 levels of
heterogeneity with $\psi^2/w^2 = 0, 0.5, 1, 2, \infty$. This yields a
grid of $4 \times5 $ different $(\psi, w)$ combinations, and we treat
every grid value as \textit{a priori} equally likely when computing
$\mathrm{ABF}_{\mathrm{av}}^{\mathrm{EE}}$. (A denser grid could be
used to obtain more precise
estimates of $\psi$ and $w$,
but this coarse grid suffices for our purposes here.)

%
\begin{table}[b]
\tabcolsep=2.7pt
\caption{Bayesian analysis of genetic
associations with recombination rate. The SNPs, estimated effect sizes
and $p$-values are taken directly from Kong et~al. (\citeyear{Kong2008}), Table~1. We\vspace*{1pt}
compute approximate BFs under the EE model using these summary
statistics. $\mathrm{ABF}^{\mathrm{EE}}_{\mathrm{single},\mathrm
{male}}$ and $\mathrm{ABF}^{\mathrm{EE}}_{\mathrm{single},\mathrm{female}}$
are approximate BFs computed using only male
subgroup and female subgroup data, respectively. $\mathrm{ABF}_{\mathrm
{fix}}^{\mathrm{EE}}$
and $\mathrm{ABF}_{\mathrm{av}}^{\mathrm{EE}}$ are approximate BFs
computed using both males
and females jointly, either assuming no heterogeneity ($\mathrm
{ABF}_{\mathrm{fix}}^{\mathrm{EE}}$) or averaging over a range of
values of heterogeneity
($\mathrm{ABF}_{\mathrm{av}}^{\mathrm{EE}}$)}\label{recombrsttbl}
{\fontsize{7pt}{10pt}\selectfont
\tabcolsep=2pt
\begin{tabular*}{\tablewidth}{@{\extracolsep{\fill}}@{}lcc c c c c c@{}}
\hline
 & \multicolumn{2}{c}{\textbf{Male}} & \multicolumn{2}{c}{\textbf{Female}} & \multicolumn{2}{c}{\textbf{Meta BFs}}\\[-6pt]
 & \multicolumn{2}{c}{\hrulefill} & \multicolumn{2}{c}{\hrulefill} & \multicolumn{2}{c}{\hrulefill}\\
\textbf{SNP} & \textbf{Effect ($\bolds{p}$-value)} & $\mathbf{ABF^{EE}_{single,\,male}}$
             & \textbf{Effect ($\bolds{p}$-value)} & $\mathbf{ABF^{EE}_{single,\,female}}$
             & $\mathbf{ABF_{fix}^{EE}} $ & $\mathbf{ABF_{av}^{EE}}$
\\
\hline
rs3796619 & $-$67.9 ($1.1\times10^{-14}$)  &\phantom{0}$10^{11.12}$& 67.6 ($7.9\times10^{-6}$) &$10^{2.81}$& $10^{3.07}$ & $10^{13.91}$ \\
rs1670533 & $-$66.1 ($1.8 \times10^{-11}$) &$10^{8.06}$& 92.8 ($4.1\times10^{-8}$) &$10^{4.55}$& $10^{1.10}$ & $10^{12.58}$ \\
rs2045065 & $-$66.2 ($1.6 \times10^{-11}$) &$10^{8.11}$& 92.2 ($6.0\times10^{-8}$) &$10^{4.40}$& $10^{1.18}$ & $10^{12.49}$ \\
\hline
\end{tabular*}}
\end{table}

The\vspace*{1pt} resulting BFs are shown in Table~\ref{recombrsttbl}.
Notably, the association signal in the joint analysis of males and
females, allowing for heterogeneity, $\mathrm{ABF}_{\mathrm
{av}}^{\mathrm{EE}}$, is many
orders of magnitude larger than either of the subgroup-specific BFs,
which are themselves larger than the BF under a fixed effects model,
$\mathrm{ABF}_{\mathrm{fix}}^{\mathrm{EE}}$.
This analysis illustrates two simple but important points. First, a
joint analysis can yield a considerably stronger signal than
subgroup-specific analyses. Second, a standard fixed-effects
meta-analysis would be ineffective in this case. Of course, in general,
the ``right'' level of heterogeneity is unknown; $\mathrm{ABF}_{\mathrm
{av}}^{\mathrm{EE}}$
deals with this by averaging over different levels of heterogeneity.
This ability to average over unknown quantities is an attractive
feature of the Bayesian approach, although it can also be helpful to
examine the components of this average separately (see the next application).

Our methods can also assess which associations show evidence for heterogeneity.
This could help identify potentially interesting interactions (as in
this case) or potentially suspect association signals (see next
example). In these data $\mathrm{ABF}_{\mathrm{av}}^{\mathrm{EE}}$
is substantially larger
than $ \mathrm{ABF}_{\mathrm{fix}}^{\mathrm{EE}} $, indicating that
the data are
inconsistent with the fixed effects assumption of equal effects in both
subgroups. This comparison is a Bayesian analogue of the frequentist
test for heterogeneity in a random effects meta-analysis.
To further investigate heterogeneity, we compare BFs for different
values of the heterogeneity parameter ($\psi^2/w^2$); in this case the
data are consistent with infinite values for this parameter
(i.e., $w^2=0$).

This example illustrates that an association analysis accounting for
heterogeneity can identify associations that would
be missed by a fixed effects analysis that ignores heterogeneity. Next
we attempt to exploit this to identify novel
associations in a large-scale genome-wide association study.

\subsection{Global Lipids genome-wide association study}\label{sec3.2}

We now return to motivating application~1,
a large scale meta-analysis of genome-wide genetic association studies
of blood lipids phenotypes conducted by the Global Lipids consortium
[\citet{Teslovich2010}]. In this study, more than 100,000 individuals
of European ancestry
were amassed through 46 separate studies (grouped into 25 studies in
their final analysis).
For each individual, measures of total cholesterol (TC), low-density
lipoprotein cholesterol (LDL-C), high-density lipoprotein cholesterol
(HDL-C) and triglycerides (TG) were obtained. Genotypes at 2.7 million
SNPs across the genome were also measured or imputed. In each study,
the phenotypes were independently quantile normal transformed; single
SNP association tests were performed for all SNPs and all phenotypes
using the linear model (\ref{studylinmodel}) and estimated effects
and their standard errors were computed. The meta-analysis combined
these summary data using the software METAL
[\citet{Willer2010}] to compute a weighted $Z$ statistic that we
discuss later [specifically they used equation (\ref
{weightedZscoreseqn}) with weights $w_s = \sqrt{n_s}$].
This can be viewed as an approximation to a fixed effects analysis
under the ES model (see Section~\ref{fixmodelsec}). \citet{Teslovich2010} reported 168 SNP-phenotype associations exceeding their
``genome-wide significant'' threshold ($p$-value $< 5 \times10^{-8}$)
and identified 95 genes, with 59 showing genome-wide significant
association for the first time.

We hypothesized that taking account of heterogeneity among subgroups
might help identify additional novel associations, so we reanalyzed the
data using our Bayesian tools to allow for heterogeneity.
We were able to obtain access to summary data from each study, in the
form of an estimated effect size (computed from the
quantile-transformed phenotype data) and its standard error for each
SNP in each study. With these data we can perform analyses under the EE
model for the quantile normalized data (rather than the ES model
effectively used in the original analysis).
We computed ABFs under increasing amounts of heterogeneity: the fixed
effects model, the ``limited heterogeneity'' CEFN model, and the maximum
heterogeneity model.
In each case we assumed a discrete uniform prior on the overall genetic
effect size [i.e., $(k^2+1)w^2$ in the CEFN model and $w^2 + \psi^2$
in the other models] on the set $\{0.1^2, 0.2^2, 0.4^2, 0.6^2,0.8^2\}$.
For the fixed effects model, $\psi=0$; for the maximum heterogeneity
model, $w=0$; and for the CEFN model, we set $k=0.326$, which gives a
prior probability of $1/1000$ that the genetic effect in each study has
an opposite sign to $\bar{\beta}$.
Although a fully automated Bayesian analysis would naturally average
over the different models
for heterogeneity, in practice, because of their different sensitivity
to particular data features
(see below), we found it helpful to examine each separately.

As an initial check on data handling we verified that our Bayesian
fixed effects analysis produced similar
results to the original fixed effects analysis. As expected, we found
that $\mathrm{ABF}_{\mathrm{fix}}^{\mathrm{EE}}$
ranked SNP associations very similarly to the original
reported (ES model) $p$-values. However, there were some notable
exceptions. In particular,
a few SNPs showed a much stronger association signal in $\mathrm
{ABF}_{\mathrm{fix}}^{\mathrm{EE}}$ than the original analysis. Further
investigation suggested that these results likely reflected the EE
analysis being less
robust to data processing errors than the original ES analysis. For
example, SNP rs17061870 with LDL phenoype had
a huge signal in our EE analysis ($\mathrm{ABF}_{\mathrm
{fix}}^{\mathrm{EE}} > 10^{17}$) and
a very modest signal in the original analysis ($p$-value${}= 0.028$),
but examination of the study-specific data for this SNP showed a
suspicious pattern: the $p$-value was $2 \times10^{-31}$ in one study
(the Family Heart Study, FHS), but no smaller than 0.1 in the 5 other
studies for which this SNP had genotype
data available. Furthermore, the very small $p$-value in FHS was driven
primarily by a very small, probably erroneous, estimate of the residual
error in that study (the sample size of this particular study is not
large) which under the EE model results in a very high weight on that
study, but in the ES model does not (see Section~\ref{fixmodelsec}).
We emphasize that we performed the EE analysis here because it was what
we were able to do easily with the available summary data, rather than
because we prefer it.

\begin{figure}

\includegraphics{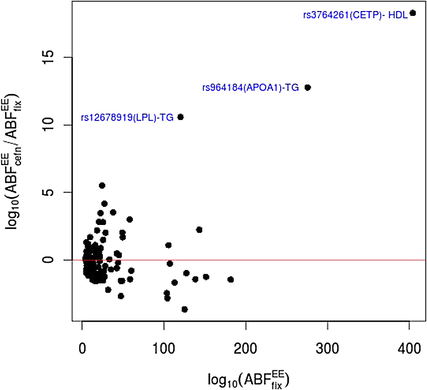}

\caption{Assessment of evidence
for heterogeneity in 168 reported phenotype-SNP association signals
from Teslovich et~al. (\citeyear{Teslovich2010}).
Large values on the $y$ axis indicate stronger support for the model,
allowing for limited heterogeneity ($\mathrm{ABF^{EE}_{cefn}}$) compared
with the model with no heterogeneity ($\mathrm{ABF^{EE}_{fix}}$).
The three highlighted points correspond to associations with
overwhelming evidence for heterogeneity, $\mathrm{ABF^{EE}_{cefn}}/
\mathrm{ABF^{EE}_{fix}}>10^{10}$; forest plots for these are shown in
Figure~\protect\ref{forestplotfigure}.}\label{cefnvsfixfigure}
\end{figure}

\begin{sidewaysfigure}

\includegraphics{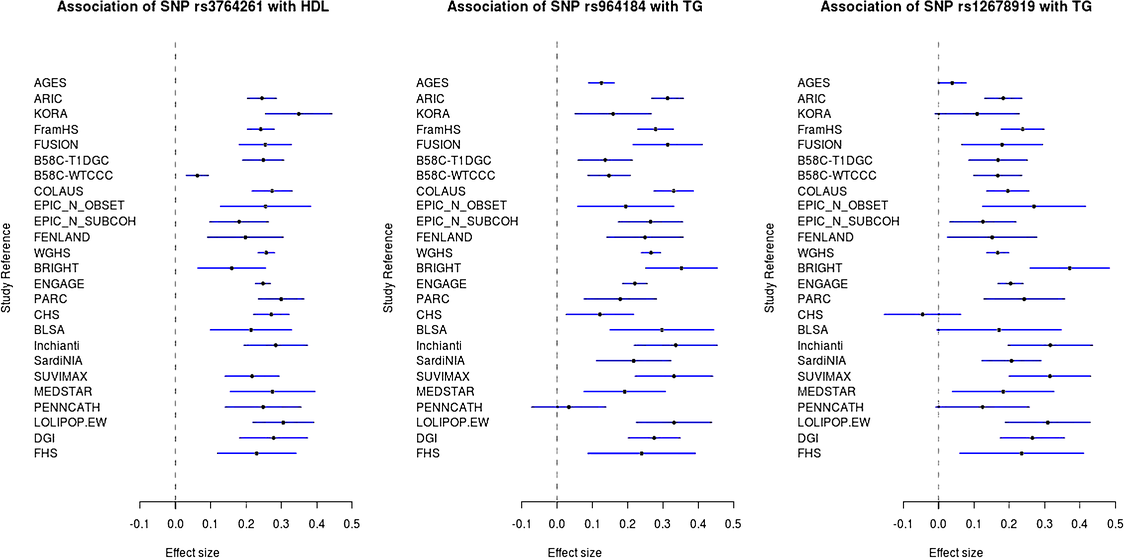}

\caption{Forest plots for the
three highlighted association signals in Figure~\protect\ref{cefnvsfixfigure}, which showed overwhelming evidence for
heterogeneity of (apparent) effects.}\label{forestplotfigure}
\end{sidewaysfigure}

Next we assessed evidence for heterogeneity of effects in the 168
association signals reported by the original analysis. We did this by
comparing the support for the limited heterogeneity model ($\mathrm
{ABF^{EE}_{cefn}}$) with the support for the no heterogeneity model
($\mathrm{ABF^{EE}_{fix}}$).
The majority of phenotype-SNP pairs ($111/168$) showed stronger support
for the no heterogeneity model (Figure~\ref{cefnvsfixfigure}).
This result was surprising to us: this meta-analysis involved a large
number of different studies, encompassing a range of different
enrollment criteria,
so we expected to find much stronger evidence for heterogeneity among
studies. We did find three strong associations that showed
overwhelming support for heterogeneity \mbox{($\mathrm
{ABF^{EE}_{cefn}}/\mathrm{ABF^{EE}_{fix}} >10^{10}$}; Figure~\ref{cefnvsfixfigure}).
Forest plots for these SNPs (Figure~\ref{forestplotfigure})\vadjust{\goodbreak} suggest
that in all three cases this signal for heterogeneity comes from modest
variation
in effect size among all studies, rather than a strong difference in
one or a few studies (although the \mbox{B58C-WTCCC} study is, arguably,
something of
an outlier at rs3764261).

%

%

Finally, we addressed the primary question of interest: whether
allowing for heterogeneity across studies yields novel
associations. To do this, we performed a genome-wide analysis for each
phenotype, in each case excluding all SNPs within 1~Mb of any SNPs
originally reported as associated with that phenotype. We searched for
SNPs that showed strong evidence for association under one of the
heterogeneity models ($\mathrm{ABF^{EE}_{cefn}}$
or $\mathrm{ABF}_{\max\mathrm{H}}^{\mathrm{EE}} \ge10^6$, where $\mathrm{ABF_{\max\mathrm{H}}^{EE}}$
denotes the approximate Bayes factor computed by restricting $w=0$, see
Section~\ref{maxhmodelsec} for precise definition) but not under the
fixed-effects model ($\mathrm{ABF^{EE}_{fix}}< 10^6$). This threshold
($10^6$) corresponds very roughly to, and is perhaps slightly more
conservative than, the threshold effectively used in the initial
analysis. (We used the same threshold for all three models for
simplicity, but different thresholds might be more appropriate; for
example, one might
prefer a more stringent threshold for $\mathrm{ABF_{\max\mathrm{H}}^{EE}}$ because
strong heterogeneity is unexpected in this context.)

%
\begin{table}
\tabcolsep=0pt
\caption{Association signals that show strong
association under the models allowing for heterogenetiy
($\mathrm{ABF^{EE}_{cefn}}$ or $\mathrm{ABF_{\max\mathrm{H}}^{EE}} \ge10^6$) but less\vspace*{1pt} strong
under a model with no heterogeneity ($\mathrm{ABF^{EE}_{fix}}< 10^6$). It
seems likely that the last two of these represent false positive
associations, but we include them in the table for completeness (see
text for discussion)}\label{lipidsmisstbl}
\begin{tabular*}{\tablewidth}{@{\extracolsep{\fill}}@{}lll d{2.1} d{2.1} c@{}}
\hline
\textbf{Phenotype} & \multicolumn{1}{c}{\textbf{SNP}} & \multicolumn{1}{c}{\textbf{Gene region}}
& \multicolumn{1}{c}{$\bolds{\log}_{\mathbf{10}} \bolds{(}\mathbf{BF^{EE}_{fix}}\bolds{)}$}
& \multicolumn{1}{c}{$\bolds{\log}_{\mathbf{10}} \bolds{(}\mathbf{BF}^{\mathbf{EE}}_{\bolds{\max}\,\mathbf{H}}\bolds{)}$}
& $\bolds{\log}_{\mathbf{10}} \bolds{(}\mathbf{BF^{EE}_{cefn}}\bolds{)}$
\\
\hline
LDL & rs1800978 & {5'UTR of ABCA1} & 5.2 & 3.4 & 6.0 \\
TG & rs1562398 & {Flanking KLF14} & 5.3 & -0.2 & 6.5 \\
HDL & rs11229165 & {Flanking OR4A16} & 4.6 & 4.9 & 6.4 \\
HDL & rs7108164 & {Flanking OR4A42P} & 4.2 & 4.9 & 6.3 \\
HDL & rs11984900 & {N.A.} &-1.1 & 16.6 & 6.2 \\
HDL & rs6995137 & {Flanking SFRP1} & -0.4 & 6.9 & 4.8 \\
\hline
\end{tabular*}
\end{table}

Overall we found 42
SNP-phenotype associations satisfying this criteria (after removing
SNPs in LD with one another), representing associations potentially
missed by the original analysis. However, detailed investigation
suggested that 36 of these were not genuine associations. Specifically,
these 36 associations, which showed strong signals in $\mathrm
{ABF_{\max\mathrm{H}}^{EE}}$ only, were driven by strong associations in the FHS
study that are likely due to data processing errors (the FHS $p$-values
at these SNPs for quantile transformed phenotypes were many orders of
magnitude smaller than for the original phenotypes). We therefore
dropped the FHS data and re-performed the association analysis.

After dropping FHS, all 6 remaining signals from the previous analysis
still satisfy our association criteria (Table~\ref{lipidsmisstbl}).
Of those, the first two listed are almost certainly genuine: the genes
ABCA1 and KLF14 are reported in \citet{Teslovich2010} as associated
with other lipid phenotypes (ABCA1 with HDL and TC; KLF14 with HDL),
but not with the phenotypes we listed in Table~\ref{lipidsmisstbl},
and both reflect associations that just missed being significant in the
original fixed effects analysis. The next two
associations may also be real: they map approximately 6~Mb apart on
chromosome 11, in a region that is densely populated with olfactory
receptor genes, and this same genetic region is also identified in a
multivariate
\mbox{association} analysis of these same data [\citet{stephens2013}],
although we know of no further independent evidence to support them.
One slight cause for caution is that, in humans, SNPs this far apart
would usually not be correlated with one another, but these two are
slightly correlated ($r^2 \approx0.07$ in
the European 1000 Genomes data), raising the possibility of mapping errors.
That is, the precise locations of these SNPs may be in question,
and certainly it is difficult to say which genes they might implicate;
the table simply lists the nearest gene for reference. Finally, based
on examination of the raw data, we suspect that the
last two associations are false positives, driven by apparent anomalies
in a single study (this time \mbox{B58C-WTCCC}).

In summary, we find that the original fixed effects analysis in \citet{Teslovich2010} was highly effective.
This may seem surprising, since these data seem
to provide ample opportunities for heterogeneity of effects. Indeed,
some of the associations identified by the original study do show a
substantially stronger signal in analyses allowing for heterogeneity
(Figure~\ref{cefnvsfixfigure}), and a genome-wide association\vadjust{\goodbreak}
analysis allowing
for heterogeneity identified at least two apparently real associations
that just missed being significant under the original fixed effects analysis.
Thus, despite the success of the fixed effects analysis,
analyses allowing for heterogeneity could modestly increase in power
for GWAS meta-analyses in general.
On the other hand, our results also provide
a cautionary tale: in the context of meta-analysis of genetic
association studies, when associations appear only under models
allowing for strong heterogeneity, and not under fixed effects models,
the reasons for the discrepancy must be examined carefully and the
results interpreted critically. Indeed, we found
that searching for SNPs showing strong heterogeneity is
an effective way to identify data processing errors that may otherwise
lurk undetected!

\subsection{Heterogeneity of eQTLs among populations}\label{sec3.3}\label{popeqtl}

Now we consider our second motivating application, examining
heterogeneity in the effects of expression quantitative trait loci
(eQTLs) among populations. An eQTL is a genetic variant (here, a SNP)
that is associated with gene expression.
Understanding heterogeneity of eQTL effects among population subgroups
is important for several reasons. For example, it is important for
designing and
interpreting experiments, because it influences how generalizable results
obtained in one subgroup are to other subgroups.
In addition, identifying heterogeneous effects could yield insights
into biological differences among subgroups: if an eQTL is more active
in one subgroup than others, it
may indicate a difference in the regulatory mechanisms operating in
that subgroup.

To assess heterogeneity of eQTLs among European, African and Asian
subgroups, we analyzed gene expression measurements from \citet{Stranger2007}, obtained using the Illumina Sentrix Human-6 Expression
BeadChip, on lymphoblastoid cell lines. Specifically,
we considered the subset of 141 cell lines [41~Europeans (CEU), 59
Asians (ASN) and 41 Africans (YRI)] that were fully sequenced in the
pilot project of the 1000 Genomes project [\citet{Durbin2010}].
We analyzed the 8427 distinct autosomal genes that were confirmed to
be expressed in the same African samples by an independent experiment
[\citet{Pickrell2010}]. We used SNP genotype data on 14.4 million SNPs
from the final release (March, 2010) of the pilot SNP calls from the
1000 genomes project, with no additional allele frequency filtering. In
addition to the original normalization \citet{Stranger2007}, we
performed quantile normal transformations to expression values for each
gene, separately within each population group, to reduce the influence
of outliers or other deviations from normality.
Previous studies have shown that most eQTLs are located near to the
gene whose expression they influence (so-called ``\textit{cis}-eQTLs'').
Therefore, for each gene we restricted our association analysis to the
``\textit{cis} SNPs'' which lie within the region 500~kb upstream of the
transcription start site and 500~kb downstream of the transcription end site.

%
\begin{table}[b]
\tablewidth=205pt
\caption{Estimated heterogeneity of eQTL effects
among Europeans, Africans and Asians, obtained by fitting a
hierarchical model to combine information across genes}\label{hmhettbl}
\begin{tabular*}{205pt}{@{\extracolsep{\fill}}@{}lcc@{}}
\hline
$\bolds{\phi^2/\omega^2}$ & \textbf{Posterior mean} & \textbf{95\% credible interval}\\
\hline
0 & 0.700 & (0.640, 0.753) \\
$1/4$ & 0.265 & (0.195, 0.330) \\
$1/2$ & 0.015 & (0.002, 0.052) \\
1 & 0.008 & (0.003, 0.016) \\
$2/1$ & 0.007 & (0.002, 0.015) \\
$4/1$ & 0.004 & (0.001, 0.012) \\
$\infty$ & 0.003 & (0.000, 0.010) \\
\hline
\end{tabular*}
\end{table}

Our analysis focuses on the question: how much do eQTL effects vary
among continental groups?
To assess this, we applied the ES model with $\sqrt{\phi^2+\omega^2}
\in\{0.1, 0.2, 0.4, 0.8, 1.6\}$ and $\phi^2/\omega^2 \in\{0, 1/4,
1/2, 1, 2, 4, \infty\}$, producing a total grid of 35 different $(\phi,\omega)$ combinations.
These values were chosen to cover a~wide range of possible effect sizes
and levels of heterogeneity. Since the amount of heterogeneity is our
main interest, we estimate the weights $(\pi)$ on the 35 combinations
using a Bayesian hierarchical model that jointly analyzes all 8427
genes. In brief, this model
assumes the following: (i) the data at each gene are independent; (ii)
each gene has at most one eQTL, with each SNP being equally likely; and
(iii) that each eQTL draws its
$(\phi,\omega)$ value from the grid of 35 different values, according
to $\pi$.
In addition, we assign a uniform prior to $\pi$. Under these
assumptions, we implement a Markov Chain Monte Carlo (MCMC) algorithm
to perform posterior inference on $\pi$. [See \citet{wenthesis},
\citet{flutre} for full details of the computational methods and
modeling assumptions.]

The results (Table~\ref{hmhettbl})
suggest that eQTL effects typically vary little among subgroups: the
estimates from the hierarchical model put 97\% of the total weight on
the two smallest heterogeneity parameters, 0 and $1/4$. (For completeness
we also include estimated grid weights on $\phi^2+\omega^2$, which
control the average eQTL effect sizes, in
Table~\ref{hmefftbl}.)

%
\begin{table}
\tablewidth=205pt
\caption{Estimated standard deviation of average
eQTL effects ($\sqrt{\omega^2 + \phi^2}$), obtained by fitting a
hierarchical model to combine information across genes}\label{hmefftbl}
\begin{tabular*}{205pt}{@{\extracolsep{\fill}}@{}lcc@{}}
\hline
$\bolds{\sqrt{\omega^2+\phi^2}}$ & \textbf{Posterior mean} & \textbf{95\% credible interval}\\
\hline
0.1 & 0.004 & (0.001, 0.008) \\
0.2 & 0.004 & (0.001, 0.007) \\
0.4 & 0.008 & (0.003, 0.014) \\
0.8 & 0.976 & (0.966, 0.983) \\
1.6 & 0.007 & (0.002, 0.015) \\
\hline
\end{tabular*}
\end{table}

The above analysis effectively assumes that each eQTL is active in all
three populations and allows for heterogeneity by allowing that the
effect size may vary among populations. That is, it is effectively a
model for ``quantitative heterogeneity''. A different model for
heterogeneity is that some eQTLs may be active in only a subset of the
populations, with no effect in others, that is, that heterogeneity
might be qualitative, with some eQTLs being
``population-specific''.\footnote{It could be objected that the notion
of a ``population-specific'' eQTL is too simplistic and that apparent
absence of effects in some populations more likely reflects very small,
nonzero, effects. While sympathetic to this argument, we also find the
simplicity has a certain appeal, and we view such models as potentially
useful nonetheless.}
Although our quantitative-heterogeneity analysis above suggests that
heterogeneity is generally\vadjust{\goodbreak} low,
it does not preclude the existence of some population-specific effects, so
we performed an additional analysis to assess how common such
population-specific effects might be. To apply our methods to this
situation, we introduce $C$ to denote a binary string of indicators for
whether an eQTL is active (i.e., has nonzero effect) in each
population. For example, $C=(110)$ would indicate that the eQTL is
active only in the first two populations. For the three populations in
our data, $C$~has $2^3$ possible values, which we refer to as
``configurations''. The support (BF) for each configuration, relative to
the null model $C=(000)$, is easily computed. For example, for $C=(110)$,
%
\begin{eqnarray}
\mathrm{BF}_{C=(110)} & =& \frac{P ( \mathbf{y}_1,\mathbf{y}_2,\mathbf{y}_3 | \mathbf{g}_1,
\mathbf{g}_2, \mathbf{g}_3,C = (110)
)}{P ( \mathbf{y}_1,\mathbf{y}_2,\mathbf{y}_3 |\mathbf{g}_1, \mathbf{g}_2,
\mathbf{g}_3, C = (000) )}
\nonumber\\[-8pt]\\[-8pt]
& =& \frac{P ( \mathbf{y}_1,\mathbf{y}_2 |
\mathbf{g}_1, \mathbf{g}_2 )}{P( \mathbf{y}_1,\mathbf{y}_2 | H_0)}.\nonumber
\end{eqnarray}
[The simplification is due to the assumption that the vectors of
residual errors in~(\ref{studylinmodel}) are independent across populations.]

%
\begin{table}
\caption{Estimated proportion of eQTLs that are
shared among the three continental subgroups, represented by CEU
(European), ASN (Asian) and YRI (African) samples. Estimates come from
fitting a hierarchical model to combine information across genes; see
text. The vast majority of eQTLs are estimates to be shared among all three subgroups}\label{hmetatbl}
\begin{tabular*}{\tablewidth}{@{\extracolsep{\fill}}@{}lcc@{}}
\hline
\textbf{Configuration} & \textbf{Estimate (posterior mean)} & \textbf{95\% credible interval}\\
\hline
CEU only & 0.001 & (0.000, 0.003) \\
ASN only & 0.001 & (0.000, 0.008) \\
YRI only & 0.001 & (0.000, 0.004) \\
CEU and YRI & 0.001 & (0.000, 0.013) \\
ASN and YRI & 0.017 & (0.010, 0.035) \\
CEU and ASN & 0.026 & (0.014, 0.041) \\
CEU and ASN and YRI & 0.953 & (0.934, 0.970)\\
\hline
\end{tabular*}
\end{table}

To estimate the proportion of eQTLs that follow each (nonnull)
configuration, we introduce hyperparameters, $\eta_{001},\ldots,\eta
_{111}$, to represent the frequency of each nonnull configuration. So
each eQTL draws its configuration $C$ independently according to $\eta
$. Furthermore, given $C$, we assume the standardized eQTL effect sizes
follow the CEFN prior with $k=0.314$ and $\omega$ drawn independently
from a grid $\{0.1, 0.2, 0.4, 0.8, 1.6\}$ according to weights $\pi'$.
Putting this all together into a single hierarchical model, with
uniform priors on $\pi'$ and $\eta$, we use MCMC to
sample from the joint posterior distribution on all parameters. Full
details are given in \citet{wenthesis}, \citet{flutre}.

\begin{figure}[t!]
\includegraphics{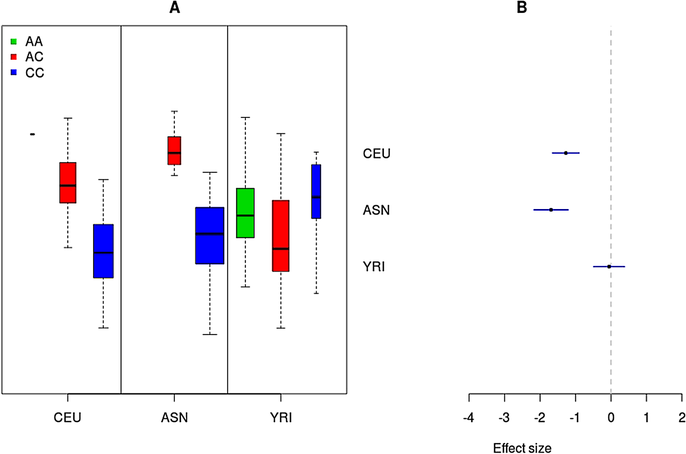}
\caption{Example of a potential
population specific eQTL presented only in CEU and ASN. \textup{A}:~Boxplots of
the gene expression levels of gene CMAH (Ensemble ID ENSG00000168405)
according to genotypes of SNP rs6906102 in the three Hapmap
populations. \textup{B}: forest plot of estimated effect sizes of this eQTL.
Allele A of this SNP has allele frequencies 0.22, 0.08 and 0.67 in ASN,
CEU and YRI, respectively.}\label{eqtlcasharefigure}
\end{figure}

\begin{figure}[t]
\includegraphics{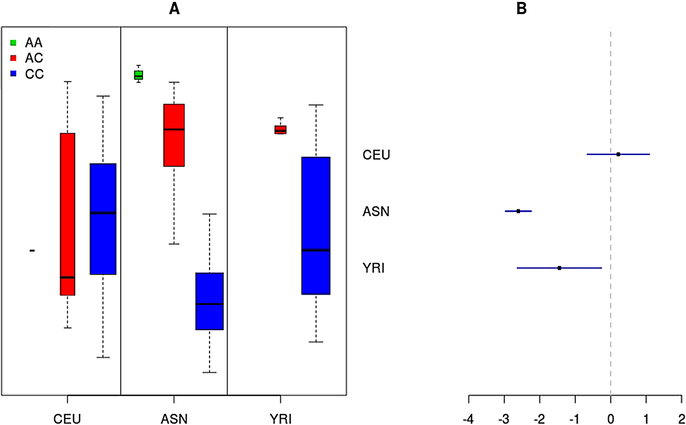}
\caption{Example of a potential
population specific eQTL presented only in ASN and YRI. \textup{A}: Boxplots of
the gene expression levels of gene PAQR8 (Ensemble IDENSG00000170915)
according to genotypes of SNP rs3180068 in the three Hapmap
populations. \textup{B}: forest plot of estimated effect sizes of this eQTL.
Allele A of this SNP has sample allele frequencies 0.14, 0.22 and 0.07
in ASN, CEU and YRI, respectively.}\label{eqtlaysharefigure}
\end{figure}

The resulting estimates for $\eta$ are given in Table~\ref{hmetatbl}. Consistent with the conclusions on low overall
heterogeneity above, the vast majority of eQTLs behave consistently
across populations: indeed, we estimate $95\%$ of eQTLs to be active in
all three populations. Nonetheless,
we find some evidence for occasional deviations from this pattern, with
approximately $2\%$ of eQTLs being active only in European and Asian
samples, and $2\%$ being active only in Asian and African. Illustrative
examples of
potential exceptions to the general rule of sharing among populations
are shown in Figures~\ref{eqtlcasharefigure} and \ref
{eqtlaysharefigure}.

%
%
%

\section{Analytic expressions for the Bayes factors and connections
with frequentist statistics}\label{sec4} \label{secanalytic}

We now provide analytic expressions for the ABFs mentioned above
(Proposition \ref{abfesprop} below). These expressions provide
intuitive insights and
highlight connections with standard frequentist test statistics,
effectively establishing the ``implicit prior assumptions'' underlying
some standard frequentist procedures. We start by introducing
necessary notation:
%
\begin{itemize}
\item\textit{Association testing in a single subgroup}. Consider
analyzing a single subgroup,~$s$. Let $\hat\beta_s$ and $\hat\sigma
_s$ denote the least square estimates of $\beta_s$ and $\sigma_s$
from the linear regression model (\ref{studylinmodel}) using only
data from subgroup $s$. The following expressions give an estimate for
the standardized effect $b_s$ ($\hat b_s$), its standard error under~$H_0$ ($\delta_s$) and a $t$-statistic for testing $b_s=0$ ($T_s$):
%
\begin{eqnarray}
\hat b_s &:=& \hat\beta_s/ \hat\sigma_s,
\\
\delta^2_s &:=& \frac{1}{\mathbf{g}_s'\mathbf{g}_s - n_s \bar g_s^2},
\\
T_s^2 &:=& \frac{\hat b_s^2}{\operatorname{se}(\hat b_s)^2} = \frac{\hat
\beta
_s^2}{\hat\sigma_s^2 \delta_s^2}.
\end{eqnarray}
Note that $T_s$ is also equal to $\hat\beta_s/ \operatorname{se}(\hat
\beta
_s)$, which is the usual t-statistic for testing $\beta_s=0$.

Both \citet{Wakefield2009} and \citet{Johnson2008} derive the following
approximate BF for testing $b_s \sim\mathrm{N}(0, \phi^2)$ vs. $ b_s
= 0 $:
%
\begin{equation}
\label{singleabf} \mathrm{ABF_{single}^{ES}}(T_s,
\delta_s; \phi):= \sqrt{\frac
{\delta
_s^2}{\delta_s^2 + \phi^2}} \exp \biggl(
\frac{T_s^2}{2} \frac{\phi
^2}{\delta^2_s + \phi^2} \biggr).
\end{equation}
As noted by Wakefield, if $\phi$ is chosen differently for each SNP,
and proportional to the value of $\delta^2_s$ for that SNP, then
$\mathrm{ABF_{single}^{ES}}$ ranks the SNPs in the same way as the
usual test
statistic $T_s$. This result connects the standard frequentist analysis
to a particular (approximate) Bayesian analysis in the case of a single
subgroup.
Proposition \ref{abfesprop} below extends this to multiple
subgroups, allowing for
heterogeneity among subgroups.

\item\textit{Testing average effect in a random effect meta-analysis
model}. Consider the standard frequentist test of $\bar b =0$ in a
random effect meta-analysis of all subgroups, with $b_s \sim\mathrm
{N}(\bar b, \phi^2)$. If $\phi$ is considered known, then an estimate
for $\bar b$, its standard error $\zeta$ and a test statistic
$\mathcal{T}_{\mathrm{ES}}
^2$ for testing $\bar b = 0$ are given by
%
\begin{eqnarray}
\hat{\bar b} &:=& \frac{\sum_s (\delta_s^2+ \phi^2)^{-1}\hat
b_s}{\sum_s (\delta_s^2+ \phi^2)^{-1}},
\\
\zeta^2 &:=& \frac{1}{\sum_s(\delta_s^2+ \phi^2)^{-1}},
\\
\mathcal{T}_{\mathrm{ES}}^2 &:=& \frac{\hat{\bar b}^2 }{\operatorname{se}(\hat{\bar b})^2}.
\end{eqnarray}
Applying Johnson's idea [\citeauthor{Johnson2005} (\citeyear{Johnson2005,Johnson2008})], we can
``translate'' this test statistic into an approximate BF for testing
$\bar b \sim\mathrm{N}(0, \omega^2)$ vs. $ \bar b = 0 $, which yields
%
\begin{equation}
\label{fixabf} \mathrm{ABF_{single}^{ES}}\bigl(
\mathcal{T}_{\mathrm{ES}}^2, \zeta; \omega\bigr):= \sqrt{
\frac
{\zeta^2} {\zeta^2 + \omega^2}}\exp \biggl( \frac{\mathcal
{T}_{\mathrm{ES}}^2}{2} \frac
{\omega^2}{\zeta^2+ \omega^2} \biggr).
\end{equation}
\end{itemize}

Now $\mathrm{ABF}^{\mathrm{ES}}(\phi,\omega)$ can be written as a simple product
of the ABFs (\ref{singleabf}) and~(\ref{fixabf}).

%
\begin{prop}
\label{abfesprop}
Under the ES model, applying a version of Laplace's method to approximate
$\mathrm{BF^{ES}}(\phi,\omega)$ yields the approximation
%
\begin{eqnarray}\label{abfes}
\mathrm{BF^{ES}}(\phi,\omega) &\approx&
\mathrm{ABF}^{\mathrm{ES}}(\phi,\omega)
\nonumber\\[-8pt]\\[-8pt]
&:=& \mathrm{ABF_{single}^{ES}}
\bigl(\mathcal {T}_{\mathrm{ES}}^2, \zeta;\omega\bigr) \cdot\prod
_s \mathrm {ABF_{single}^{ES}}
\bigl(T_s^2, \delta_s; \phi\bigr)\nonumber
\end{eqnarray}
and $\mathrm{ABF}^{\mathrm{ES}}(\phi,\omega)$ converges (almost
surely) to $\mathrm{BF^{ES}}(\phi,\omega)$ as $n_s \to\infty$ for
all subgroups $s$.
\end{prop}

\begin{pf}
See Appendix A.1 of the supplementary material [\citet{suppA}].
\end{pf}

%
\begin{note}
If the study-specific residual error variances, $\sigma_s$, are
considered known (rather than being assigned a prior distribution) and
used in place of~$\hat{\sigma}_s$ to compute $\mathrm{ABF}^{\mathrm
{ES}}$, then the
approximation is exact, and $\mathrm{ABF}^{\mathrm{ES}}(\phi,\omega
)=\mathrm{BF^{ES}}(\phi,\omega)$. This fact, together with the fact that the estimators $\hat
{\sigma}_s$
are consistent for $\sigma_s$, explains, intuitively, why the
proposition holds.
\end{note}

%
\begin{note}
The numerical accuracy of $\mathrm{ABF}^{\mathrm{ES}}$ as an
approximation to $\mathrm{BF^{ES}}$ depends on sample sizes, and for
small sample sizes it may be
too inaccurate for routine application. However, a simple modification,
described in Appendix~\textup{C} of the supplementary material [Wen and Stephens (\citeyear{suppA})], yields much greater accuracy.
\end{note}

Proposition \ref{abfesprop} partitions the evidence for association
into two parts: one part reflects the evidence in each subgroup (the
second term) and the other reflects consistency of effects among
subgroups (the first term). In particular, if all subgroups show
effects in the same direction, then the first term may be large ($\gg
1$) and ``boost'' the evidence for association. A similar result holds
for the EE model (Appendix~A.2 of the supplementary material [\citet{suppA}]).

\subsection{Properties of Bayes factors}\label{sec4.1}

\subsubsection{Induced single study Bayes factors}\label{sec4.1.1}

For the ES model, in the special case of one subgroup ($S=1$), both the
actual BF and our approximations reduce to results from previous work.
Specifically, the approximation $\widehat{\mathrm{BF}}{}^{\mathrm
{ES}}$ becomes exact in this case,
and equal to the BF derived by \citet{servinstephens}, whereas
$\mathrm{ABF}^{\mathrm{ES}}$ is equal to the ABF in \citet{Wakefield2009} [see
also \citeauthor{Johnson2005} (\citeyear{Johnson2005,Johnson2008})].

\subsubsection{Noninformative subgroup data}\label{sec4.1.2}

Suppose that in one subgroup, $s$, sample genotypes vary very little.
This might arise, for example, in cross-population genetic studies, if
one SNP allele is very rare in one population. Intuitively, subgroup
$s$ then contains little information for testing $H_0$. Indeed, this
will be reflected in the standard error for the effect size, $\delta
_s$, being large, which will result in study~$s$ contributing little to
the BF. Specifically, in the limit $\delta_s \rightarrow\infty$, the
ABF (\ref{abfes}) is unaffected by the association data in study $s$
($T_s$); a similar result holds for the exact BF (Appendix~A of the supplementary material [\citet{suppA}]) and for both EE and ES.
Thus, the BF correctly reflects the noninformativeness of the data from
study $s$. Although one might expect
every reasonable statistical procedure to possess this very intuitive
property, many widely used methods do not
(e.g., Fisher's combined probability test).

\subsection{Extreme models and connections with frequentist tests}\label{sec4.2}

The proposed models are very flexible, covering a wide range of types
and degrees of heterogeneity by setting different values for $(\phi,
\omega)$ (or $k$ in the CEFN prior). Here we discuss the two extremes
of no heterogeneity (``fixed effects'') and maximum heterogeneity, and
establish connections with frequentist testing approaches in these settings.

\subsubsection{The fixed effects model}\label{sec4.2.1}\label{fixmodelsec}

The fixed effects model assumes genetic effects to be homogeneous
across subgroups, and corresponds to
$\phi=0$ in ES or $\psi=0$ in~EE.
In these cases the test statistics $\mathcal{T}_{\mathrm{ES}}$ and
$\mathcal{T}_{\mathrm{EE}}$ have
particularly simple forms, being a weighted sum of individual $T_s$
statistics from each study (often referred to as a weighted sum of $Z$
scores when sample sizes are large). Specifically,
%
\begin{equation}
\label{weightedZscoreseqn} \mathcal{T} = \frac{\sum_s w_s T_s}{\sqrt{\sum_{s'} w_{s'}^2}},
\end{equation}
where,
\begin{enumerate}[2.]
\item[1.] For the ES model, $w_s = \operatorname{se}(\hat{b}_s)^{-1} \approx
\sqrt {2n_s f_s (1-f_s)}$,\vspace*{2pt}

\item[2.] For the EE model, $w_s = \operatorname{se}(\hat{\beta}_s)^{-1}
\approx
\hat{\sigma}_s^{-1} \sqrt{2n_s f_s (1-f_s)}$,
\end{enumerate}
and $f_s$ denotes the allele frequency of the target SNP in subgroup
$s$. (The approximations come from assuming Hardy--Weinberg equilibrium
in each subgroup.) These representations clarify a key practical
difference between the ES and EE models: EE upweights studies with
small residual error variance.
Note also that $T_s$ is the same for both EE and ES, and independent of
measurement scale,
but $\sigma_s$ depends on measurement scale, so $\mathcal{T}_{\mathrm
{ES}}$ is robust to
studies using different measurement
scales (or different transformations of the phenotypes) but $\mathcal
{T}_{\mathrm{EE}}$ is not.
In addition, these representations clarify the connection between these
statistics and the methods used in
the meta-analysis software METAL [\citet{Willer2010}]. Specifically,
METAL implements tests using the weighted statistic (\ref
{weightedZscoreseqn}) with two different weighting schemes,
one corresponding to the EE model weights above and the other with the
weights equal to $\sqrt{n_s}$. This latter scheme corresponds to the
ES model only if $f_s$ is equal across studies. [Where $f_s$ varies
across studies the
weighting in the ES model seems, to us, preferable to the METAL scheme
since studies with small $f_s(1-f_s)$ provide less information.]

Returning now to the BFs, when $\phi=0$ the
ABF (\ref{abfes}) simplifies to
%
\begin{equation}
\label{abffixes} \qquad\mathrm{ABF}_{\mathrm{fix}}^{\mathrm{ES}}(\omega):= \mathrm
{ABF^{ES}}(\phi=0,\omega) = \sqrt{\frac
{\zeta^2} {\zeta^2 + \omega^2}}\cdot\exp \biggl(
\frac{\mathcal
{T}_{\mathrm{ES}}^2}{2} \frac{\omega^2}{\zeta^2+ \omega^2} \biggr),
\end{equation}
where
%
\begin{equation}
\zeta^2 = \frac{1}{\sum_s \delta_s^{-2} }.
\end{equation}
A similar expression holds for $\mathrm{ABF^{EE}_{fix}}(w):=\mathrm
{ABF^{EE}}(\psi=0,w)$.

We now answer the following question: under what prior assumptions will
$\mathrm{ABF}_{\mathrm{fix}}^{\mathrm{ES}}$ produce the
same SNP rankings as the frequentist test statistic $\mathcal
{T}_{\mathrm{ES}}$? \citet{Wakefield2009} names this kind
of prior the ``implicit $p$-value prior'', as it identifies the implicit
prior assumptions being made
when one ranks SNPs by their $p$-value computed from $\mathcal
{T}_{\mathrm{ES}}$.

Although for a \textit{given} SNP $\mathrm{ABF^{ES}_{fix}}(\omega)$ is a
monotone function of $\mathcal{T}_{\mathrm{ES}}$,
for a fixed value of $\omega$ the two statistics will not generally
rank SNPs in the same way because $\zeta$ varies among SNPs. If,
however, $\omega$ is assumed to vary among SNPs in a particular way,
then the two statistics
produce the same ranking. (A similar result holds for $\mathrm
{ABF^{EE}_{fix}}$.)
%
%
\begin{prop}[(Implicit $p$-value prior, fixed effects)]\label{propimpfixprior}
In the ES model, if the prior hyperparameter $\omega$ is allowed to
vary among SNPs, with
%
\begin{equation}
\label{impfixprior} \omega_p = K \zeta_p,
\end{equation}
where $p$ indexes SNPs and $K$ is any positive constant, then
$\mathrm{ABF}_{\mathrm{fix}}^{\mathrm{ES}}$ and $\mathcal
{T}_{\mathrm{ES}}$ will produce the same ranking of SNPs.
\end{prop}
\begin{pf}
This follows directly from substituting (\ref{impfixprior}) into
(\ref{abffixes}).
\end{pf}

%
\begin{note}
Recall that $\zeta_p$ is the standard error of $\hat{\bar b}$ for SNP
$p$, so large $\zeta_p$~corresponds to less information about ${\bar b}$ (which could occur,
for example, due to the
SNP having small minor allele frequency or being typed in only a few studies).
Recall also that large values of
$\omega_p$ correspond to a prior assumption that the effect size $\bar
b$ at SNP $p$ is likely to be large
(in absolute value). Thus, the implicit $p$-value makes the curious
assumption that
SNPs with less information have larger effects [see also Guan and Stephens (\citeyear{guanstephens})].
\end{note}

%
\begin{note}
When data on all SNPs are available on all subgroups,
and the subgroups also have similar allele frequencies at every SNP (as
might happen if
the subgroups come from a single random mating population), then
the sample genotype variance of SNP $p$ in subgroup $s$ can be well
approximated by
\mbox{$2 n_s f_p (1-f_p)$}, where $f_p$ is the population allele frequency of
SNP $p$. (Note the slight abuse of notation, since we previously
indexed $f$ by subgroup, whereas here it is indexed by~SNP.)
Consequently, the implicit frequentist prior (\ref{impfixprior}) can
be written as
%
\begin{equation}
\label{imppriorappx} \omega_p = K \sqrt{\frac{1}{f_p(1-f_p)}
\frac{1}{\sum_s{n_s}}},
\end{equation}
which is effectively the same as the single subgroup case discussed by
Wakefield (\citeyear{Wakefield2009}).
\end{note}

\subsubsection{Maximum heterogeneity model}\label{sec4.2.2}\label{maxhmodelsec}

Now consider the other end of the spectrum: models with very high heterogeneity.
Specifically, within the class of ES models with some fixed prior
expected marginal effect size ($\phi^2 + \omega^2$), the model with
$\omega= 0$ has maximum heterogeneity.
In this case, the average effect $\bar b$ is identically 0, and the
effects $b_s | \phi$ are independent, $\sim N(0,\phi^2)$.

It can be shown from (A.13) and (A.27) in Appendix~A of the supplementary material [\citet{suppA}]
that for both EE and ES, the exact BF under this setting, $\mathrm
{BF_{\max\mathrm{H}}}$, is the product of the individual BFs,
%
\begin{equation}
\label{bfprod} \mathrm{BF_{\max\mathrm{H}}} = \prod_s
\mathrm{BF}_{\mathrm{single}, s},
\end{equation}
where $\mathrm{BF}_{\mathrm{single}, s}$ is the exact BF calculated using
data only from subgroup $s$.
This relationship also holds for the ABF, that is,
%
\begin{equation}
\label{abfprodes}
\qquad\mathrm{ABF^{ES}_{\max\mathrm{H}}}(\phi):=
\mathrm{ABF^{ES}}(\phi,\omega=0) = \prod_s
\sqrt{\frac{\delta_s^2}{\delta_s^2 + \phi^2}} \exp \biggl( \frac{T_s^2}{2}
\frac{\phi^2}{\delta^2_s + \phi^2} \biggr).
\end{equation}

The frequentist test that corresponds to this ``maximum heterogeneity''
BF turns out to be the likelihood ratio test
of $H_0\dvtx  b_s=0$ (for all $s$) vs the general unconstrained alternative,
which can be written
%
\begin{equation}
\mathrm{LR_{\max\mathrm{H}} } = \prod_s
\mathrm{LR}_s,
\end{equation}
where $\mathrm{LR}_s$ is the likelihood ratio test statistic for $H_0\dvtx
b_s=0$ vs $H_1\dvtx b_s$ unconstrained. For sufficiently large samples,
$\mathrm{LR}_s$ is well approximated by
%
\begin{equation}
\lim_{n_s \to\infty} \mathrm{LR}_s \approx\exp \biggl({-
\frac
{T_s^2}{2}} \biggr),
\end{equation}
so
%
\begin{equation}
\label{llrmaxh} \mathrm{LR_{\max\mathrm{H}} } \approx\exp \biggl({-{
\frac{\sum_s
T_s^2}{2}}} \biggr).
\end{equation}
Thus, the likelihood ratio test is approximately the same as a test
based on $\sum_s T_s^2$, which (again
assuming large sample sizes) is the sum of the squared $Z$ values, and
$p$-values can be obtained by noting that under the global null
hypothesis this sum will be $\sim\chi^2_S$.
This is very similar to Fisher's approach to combining test statistics
from multiple studies.

Under what prior assumptions will $\mathrm{ABF^{ES}_{\max\mathrm{H}} }$
give the same SNP ranking as~$\sum_s T_s^2$? Under the ES model no
single $\phi$ value will give this result. However, we have the following:
%
%
\begin{prop}[(Implicit $p$-value prior, maximum heterogeneity)]\label{propimpmaxhprior}
In the ES model, if the prior hyperparameter $\phi$ is allowed to vary
among subgroups, with
%
\begin{equation}
\label{impmaxhprior} \phi^2_s = K \delta_s^2,
\end{equation}
where $K$ is a constant for all subgroups and all SNPs tested,
then $\mathrm{ABF^{ES}_{\max\mathrm{H}} }$ yields the same SNP ranking as $\sum_s T_s^2$.
\end{prop}
\begin{pf}
This follows directly from substituting (\ref{impmaxhprior}) into
(\ref{abfprodes}).
\end{pf}

%
\begin{note}
Recalling that $\delta_s$ is the standard error for $b_s$, the
implicit \mbox{$p$-}value prior (\ref{impmaxhprior}) assumes bigger effects
in subgroups with less information. There seems to be no good
justification, in general, for this prior assumption.
\end{note}

\section{Discussion}\label{sec5}

Motivated by the need to allow for heterogeneity of effects in genetic
association studies, we developed and applied a flexible toolbox of
Bayesian methods for this problem.
Our applications demonstrate how these tools can (i) identify
associations allowing for different amounts and types of heterogeneity,
and (ii) assess the amount and type of heterogeneity. The tools are
sufficiently flexible to tackle
a wide range of applications, from those involving limited
heterogeneity (e.g., a typical meta-analyses) to the more extreme
heterogeneity that might be encountered
in gene--environment interaction studies. We presented computational
methods that are
practical for large studies and highlighted connections between BFs and
standard frequentist test statistics in this context (Propositions~\ref{propimpfixprior} and \ref{propimpmaxhprior}).

The models and priors considered here are closely connected to other models
employed in meta-analysis. In particular, they are similar to mixed
effect meta-analysis models in standard frequentist approaches for
quantitative phenotypes, where the subgroup-specific intercept terms
$\mu_s$ in (\ref{studylinmodel}) are regarded as fixed effects
terms and genetic effects $\beta_s$ (or $b_s$) are regarded as random
effect terms.
Our models are also connected with, but differ in an important way
from, models used in gene--environment (G${}\times{}$E) interaction studies:
%
\begin{equation}
\label{GxElinmodel} y_i = \mu+ \beta_{e,s_i} +
\beta_{g} g_i + \beta_{{[g\dvtx e]},s_i} g_i +
e_i,\qquad e_i \sim\mathrm{N}\bigl(0, \sigma^2
\bigr),
\end{equation}
where $s_i$ denotes the subgroup membership of individual $i$, and
$\beta_{[g\dvtx e]}$ denotes the subgroup-genotype interaction terms. This
linear model can be rewritten as
%
\begin{equation}
\label{GxElinmodel2} y_i = (\mu+ \beta_{e,s_i}) + (
\beta_{g} + \beta_{{[g\dvtx e]},s_i}) g_i +
e_i,\qquad e_i \sim\mathrm{N}\bigl(0, \sigma^2
\bigr),
\end{equation}
to emphasize that each subgroup has its own intercept, $\mu+ \beta
_{e,s_i}$, and its own genetic effect, $\beta_{e} + \beta
_{{[g\dvtx e]},s_i}$. (If no marginal effect of subgroup is included, the
model makes the stronger assumption of equal intercepts for different
subgroups, which can be dangerous in practice and may lead to Simpson's
paradox [\citet{Bravata2001}].) The key difference between this model
(\ref{GxElinmodel2}) and ours (\ref{studylinmodel}) is their
assumptions on the error variances: our model allows for a different
variance in each subgroup,
whereas (\ref{GxElinmodel}) assumes them to be equal. Allowing for
subgroup-specific variances improves robustness and can improve power
[\citet{flutre}].

One important issue that we have largely ignored is the question of how
to weigh evidence
of heterogeneity in the data (e.g., large BFs for high heterogeneity
models) against an \textit{a priori}
belief that, in general, strong heterogeneity might be rare. In
principle, this is straightforward: given a prior
distribution on different types of heterogeneity, it is trivial to use
the BFs to compute posterior
distributions. However, there remains an issue of choice of appropriate priors
(which also arises in a disguised form in frequentist approaches, for
example, in selecting appropriate \mbox{$p$-}value thresholds when testing
for heterogeneity). Here we have often used
discrete uniform distributions for convenience. In general, one might
want to change this, and appropriate priors may be context-dependent.
For example, in a meta-analysis one might upweight models with limited
heterogeneity (the CEFN prior), whereas in an gene--environment
interaction study one might allow more heterogeneity.

\section*{Acknowledgment}
We thank Yongtao Guan, Tanya Teslovich, Daniel Gaffney, Michael Stein
and Peter McCullagh for helpful discussions. We thank the Global Lipids
Consortium for access to their summary data.

\begin{supplement}
\stitle{Appendix}
\slink[doi]{10.1214/13-AOAS695SUPP} 
\sdatatype{.pdf}
\sfilename{aoas695\_supp.pdf}
\sdescription{Appendices referenced in Sections~\ref{sec2}, \ref{sec2.3.1}, \ref{sec2.3.3}, \ref{sec4}
and \ref{sec4.1.2} are provided in the supplementary appendix file.}
\end{supplement}

%

\printaddresses


\begin{thebibliography}{34}
\bibitem[\protect\citeauthoryear{Bravata and Olkin}{2001}]{Bravata2001}
%
\begin{barticle}[author]
\bauthor{\bsnm{Bravata},~\bfnm{D.~M.}\binits{D.}} \AND
\bauthor{\bsnm{Olkin},~\bfnm{I.}\binits{I.}}
(\byear{2001}).
\btitle{Simple pooling versus combining in meta-analysis}.
\bjournal{Eval. Health Prof.}
\bvolume{24}
\bpages{218--230}.
\end{barticle}
%
\bptok{imsref}%
\endbibitem

\bibitem[\protect\citeauthoryear{Brown, Mangravite and
Engelhardt}{2012}]{casey}
%
\begin{bmisc}[author]
\bauthor{\bsnm{Brown}, \bfnm{C.}\binits{C.}},
\bauthor{\bsnm{Mangravite}, \bfnm{L.~M.}\binits{L.~M.}} \AND
\bauthor{\bsnm{Engelhardt}, \bfnm{B.~E.}\binits{B.~E.}}
(\byear{2012}).
\bhowpublished{Integrative modeling of eQTLs and cis-regulatory
elements suggest mechanisms
underlying cell type specifcity of eQTLs. Preprint. Available at
\arxivurl{arXiv:1210.3294}.}
\end{bmisc}
%
\bptok{imsref}%
\endbibitem

\bibitem[\protect\citeauthoryear{Burgess et~al.}{2010}]{Burgess2010}
%
\begin{barticle}[author]
\bauthor{\bsnm{Burgess},~\bfnm{Stephen}\binits{S.}},
\bauthor{\bsnm{Thompson},~\bfnm{Simon~G.}\binits{S.~G.}} \AND
\bauthor{\bsnm{Andrews},~\bfnm{G}\binits{G.}} \betal{et~al.}
(\byear{2010}).
\btitle{Bayesian methods for meta-analysis of causal relationships
estimated using genetic instrumental variables}.
\bjournal{Stat. Med.}
\bvolume{29}
\bpages{1298--1311}.
\end{barticle}
%
\bptok{imsref}%
\endbibitem

\bibitem[\protect\citeauthoryear{Butler and Wood}{2002}]{Butler2002}
%
\begin{barticle}[mr]
\bauthor{\bsnm{Butler},~\bfnm{Ronald~W.}\binits{R.~W.}} \AND
\bauthor{\bsnm{Wood},~\bfnm{Andrew~T.~A.}\binits{A.~T.~A.}}
(\byear{2002}).
\btitle{Laplace approximations for hypergeometric functions with
matrix argument}.
\bjournal{Ann. Statist.}
\bvolume{30}
\bpages{1155--1177}.
\bid{doi={10.1214/aos/1031689021}, issn={0090-5364}, mr={1926172}}
\end{barticle}
%
\bptok{imsref}%
\endbibitem

\bibitem[\protect\citeauthoryear{De Iorio et al.}{2011}]{DeIorio2011}
%
\begin{barticle}[author]
\bauthor{\bparticle{De} \bsnm{Iorio}, \bfnm{Maria}\binits{M.}},
\bauthor{\bsnm{Newcombe}, \bfnm{Paul J.}\binits{P.~J.}},
\bauthor{\bsnm{Tachmazidou}, \bfnm{Ioanna}\binits{I.}},
\bauthor{\bsnm{Verzilli}, \bfnm{Claudio J.}\binits{C.~J.}} \AND
\bauthor{\bsnm{Whittaker}, \bfnm{John C.}\binits{J.~C.}}
(\byear{2011}).
\btitle{Bayesian semiparametric meta-analysis for genetic association studies}.
\bjournal{Genet. Epidemiol.}
\bvolume{35}
\bpages{333--340}.
\bid{doi={10.1002/gepi.20581}, issn={1098-2272}, pmid={21400586}}
\end{barticle}
%
\bptok{imsref}%
\endbibitem

\bibitem[\protect\citeauthoryear{Dimas et~al.}{2009}]{Dimas2009a}
%
\begin{barticle}[author]
\bauthor{\bsnm{Dimas}, \bfnm{A.~S.}\binits{A.~S.}},
\bauthor{\bsnm{Deutsch}, \bfnm{S.}\binits{S.}},
\bauthor{\bsnm{Stranger}, \bfnm{B.~E.}\binits{B.~E.}},
\bauthor{\bsnm{Montgomery}, \bfnm{S.~B.}\binits{S.~B.}},
\bauthor{\bsnm{Borel}, \bfnm{C.}\binits{C.}} \betal{et al.}
(\byear{2009}).
\btitle{Common regulatory variation impacts gene expression in a cell
type-dependent manner.}
\bjournal{Science}
\bvolume{325}
\bpages{1246--1250}.
\end{barticle}
%
\bptok{imsref}%
\endbibitem

\bibitem[\protect\citeauthoryear{DuMouchel and Harris}{1983}]{DuMouchel1983}
%
\begin{barticle}[mr]
\bauthor{\bsnm{DuMouchel},~\bfnm{William~H.}\binits{W.~H.}} \AND
\bauthor{\bsnm{Harris},~\bfnm{Jeffrey~E.}\binits{J.~E.}}
(\byear{1983}).
\btitle{Bayes methods for combining the results of cancer studies in
humans and other species}.
\bjournal{J. Amer. Statist. Assoc.}
\bvolume{78}
\bpages{293--315}.
\bid{issn={0162-1459}, mr={0711105}}
\end{barticle}
%
\bptok{imsref}%
\endbibitem

\bibitem[\protect\citeauthoryear{Durbin et~al.}{2010}]{Durbin2010}
%
\begin{barticle}[author]
\bauthor{\bsnm{Durbin}, \bfnm{R.~M.}\binits{R.~M.}},
\bauthor{\bsnm{Altshuler}, \bfnm{D.~L.}\binits{D.~L.}},
\bauthor{\bsnm{Abecasis}, \bfnm{G.~R.}\binits{G.~R.}},
\bauthor{\bsnm{Bentley}, \bfnm{D.~R.}\binits{D.~R.}},
\bauthor{\bsnm{Chakravarti}, \bfnm{A.}\binits{A.}} \betal{et al.}
(\byear{2010}).
\btitle{A map of human genome variation from population-scale sequencing}.
\bjournal{Nature}
\bvolume{467}
\bpages{1061--1073}.
\end{barticle}
%
\bptok{imsref}%
\endbibitem

\bibitem[\protect\citeauthoryear{Eddy, Hasselblad and
Schachter}{1990}]{Eddy1990}
%
\begin{barticle}[author]
\bauthor{\bsnm{Eddy}, \bfnm{D.~M.}\binits{D.~M.}},
\bauthor{\bsnm{Hasselblad}, \bfnm{V.}\binits{V.}} \AND
\bauthor{\bsnm{Schachter}, \bfnm{R.}\binits{R.}}
(\byear{1990}).
\btitle{A Bayesian method for synthesizing evidence}.
\bjournal{International Journal of Technical Assistance in Health Care}
\bvolume{6}
\bpages{31--55}.
\end{barticle}
%
\bptok{imsref}%
\endbibitem

\bibitem[\protect\citeauthoryear{Fledel-Alon et~al.}{2011}]{Fledel-Alon2011}
%
\begin{barticle}[author]
\bauthor{\bsnm{Fledel-Alon}, \bfnm{A.}\binits{A.}},
\bauthor{\bsnm{Leffler}, \bfnm{E.~M.}\binits{E.~M.}},
\bauthor{\bsnm{Guan}, \bfnm{Y.}\binits{Y.}},
\bauthor{\bsnm{Stephens}, \bfnm{M.}\binits{M.}},
\bauthor{\bsnm{Coop}, \bfnm{G.}\binits{G.}} \betal{et al.}
(\byear{2011}).
\btitle{Variation in human recombination rates and its genetic determinants.}
\bjournal{PloS One}
\bvolume{6}
\bpages{e20321}.
\end{barticle}
%
\bptok{imsref}%
\endbibitem

\bibitem[\protect\citeauthoryear{Flutre et~al.}{2013}]{flutre}
%
\begin{barticle}[author]
\bauthor{\bsnm{Flutre}, \bfnm{T.}\binits{T.}},
\bauthor{\bsnm{Wen}, \bfnm{X.}\binits{X.}},
\bauthor{\bsnm{Pritchard}, \bfnm{J.~K.}\binits{J.~K.}} \AND
\bauthor{\bsnm{Stephens}, \bfnm{M.}\binits{M.}}
(\byear{2013}).
\btitle{A statistical framework for joint eQTL analysis in multiple tissues}.
\bjournal{PLoS Genetics}
\bvolume{9}
\bpages{e1003486}.
\end{barticle}
%
\bptok{imsref}%
\endbibitem

\bibitem[\protect\citeauthoryear{Gilad, Rifkin and
Pritchard}{2008}]{yoavjonathan}
%
\begin{barticle}[author]
\bauthor{\bsnm{Gilad}, \bfnm{Yoav}\binits{Y.}},
\bauthor{\bsnm{Rifkin}, \bfnm{Scott A.}\binits{S.~A.}} \AND
\bauthor{\bsnm{Pritchard}, \bfnm{Jonathan K.}\binits{J.~K.}}
(\byear{2008}).
\btitle{Revealing the architecture of gene regulation: The promise of
eQTL studies}.
\bjournal{Trends Genet.}
\bvolume{24}
\bpages{408--415}.
\bid{doi={10.1016/j.tig.2008.06.001}, issn={0168-9525},
mid={NIHMS71884}, pii={S0168-9525(08)00177-7}, pmcid={2583071},
pmid={18597885}}
\end{barticle}
%
\bptok{imsref}%
\endbibitem

\bibitem[\protect\citeauthoryear{Givens, Smith and
Tweedie}{1997}]{Givens1997}
%
\begin{barticle}[author]
\bauthor{\bsnm{Givens}, \bfnm{G.~H.}\binits{G.~H.}},
\bauthor{\bsnm{Smith}, \bfnm{D.~D.}\binits{D.~D.}} \AND
\bauthor{\bsnm{Tweedie}, \bfnm{R.~L.}\binits{R.~L.}}
(\byear{1997}).
\btitle{Publication bias in meta-analysis:
A~Bayesian data-augmentation approach to account for issues exemplified in the passive smoking debate.}
\bjournal{Statist. Sci.}
\bvolume{12}
\bpages{221--250}.
\end{barticle}
%
\bptok{imsref}%
\endbibitem

\bibitem[\protect\citeauthoryear{Guan and Stephens}{2008}]{guanstephens}
%
\begin{barticle}[pbm]
\bauthor{\bsnm{Guan},~\bfnm{Yongtao}\binits{Y.}} \AND
\bauthor{\bsnm{Stephens},~\bfnm{Matthew}\binits{M.}}
(\byear{2008}).
\btitle{Practical issues in imputation-based association mapping}.
\bjournal{PLoS Genetics}
\bvolume{4}
\bpages{e1000279}.
\bid{doi={10.1371/journal.pgen.1000279}, issn={1553-7404},
pmcid={2585794}, pmid={19057666}}
\end{barticle}
%
\bptok{imsref}%
\endbibitem

\bibitem[\protect\citeauthoryear{Han and Eskin}{2011}]{Han2011}
%
\begin{barticle}[pbm]
\bauthor{\bsnm{Han},~\bfnm{Buhm}\binits{B.}} \AND
\bauthor{\bsnm{Eskin},~\bfnm{Eleazar}\binits{E.}}
(\byear{2011}).
\btitle{Random-effects model aimed at discovering associations in
meta-analysis of genome-wide association studies}.
\bjournal{Am. J. Hum. Genet.}
\bvolume{88}
\bpages{586--598}.
\bid{doi={10.1016/j.ajhg.2011.04.014}, issn={1537-6605},
pii={S0002-9297(11)00155-8}, pmcid={3146723}, pmid={21565292}}
\end{barticle}
%
\bptok{imsref}%
\endbibitem

\bibitem[\protect\citeauthoryear{Johnson}{2005}]{Johnson2005}
%
\begin{barticle}[mr]
\bauthor{\bsnm{Johnson},~\bfnm{Valen~E.}\binits{V.~E.}}
(\byear{2005}).
\btitle{Bayes factors based on test statistics}.
\bjournal{J. R. Stat. Soc. Ser. B Stat. Methodol.}
\bvolume{67}
\bpages{689--701}.
\bid{doi={10.1111/j.1467-9868.2005.00521.x}, issn={1369-7412}, mr={2210687}}
\end{barticle}
%
\bptok{imsref}%
\endbibitem

\bibitem[\protect\citeauthoryear{Johnson}{2008}]{Johnson2008}
%
\begin{barticle}[mr]
\bauthor{\bsnm{Johnson},~\bfnm{Valen~E.}\binits{V.~E.}}
(\byear{2008}).
\btitle{Properties of {B}ayes factors based on test statistics}.
\bjournal{Scand. J. Stat.}
\bvolume{35}
\bpages{354--368}.
\bid{doi={10.1111/j.1467-9469.2007.00576.x}, issn={0303-6898}, mr={2418746}}
\end{barticle}
%
\bptok{imsref}%
\endbibitem

\bibitem[\protect\citeauthoryear{Kong et~al.}{2008}]{Kong2008}
%
\begin{barticle}[author]
\bauthor{\bsnm{Kong}, \bfnm{A.}\binits{A.}},
\bauthor{\bsnm{Thorleifsson}, \bfnm{G.}\binits{G.}},
\bauthor{\bsnm{Stefansson}, \bfnm{H.}\binits{H.}},
\bauthor{\bsnm{Masson}, \bfnm{G.}\binits{G.}} \betal{et al.}
(\byear{2008}).
\btitle{Sequence variants in the RNF212 gene associate with
genome-wide recombination rate}.
\bjournal{Science}
\bvolume{319}
\bpages{1398--1401}.
\end{barticle}
%
\bptok{imsref}%
\endbibitem

\bibitem[\protect\citeauthoryear{Lebrec, Stijnen and van
Houwelingen}{2010}]{Lebrec2010}
%
\begin{barticle}[mr]
\bauthor{\bsnm{Lebrec},~\bfnm{Jeremie~J.}\binits{J.~J.}},
\bauthor{\bsnm{Stijnen},~\bfnm{Theo}\binits{T.}} \AND
\bauthor{\bparticle{van} \bsnm{Houwelingen},~\bfnm{Hans~C.}\binits{H.~C.}}
(\byear{2010}).
\btitle{Dealing with heterogeneity between cohorts in genomewide SNP association studies
dealing with heterogeneity between cohorts in genomewide SNP
association studies}.
\bjournal{Stat. Appl. Genet. Mol. Biol.}
\bvolume{9}
\bpages{Art. 8, 22 pp.}
\bid{mr={2594947}}
\end{barticle}
%
\bptok{imsref}%
\endbibitem

\bibitem[\protect\citeauthoryear{Li and Begg}{1994}]{Li1994}
%
\begin{barticle}[mr]
\bauthor{\bsnm{Li},~\bfnm{Zhaohai}\binits{Z.}} \AND
\bauthor{\bsnm{Begg},~\bfnm{Colin~B.}\binits{C.~B.}}
(\byear{1994}).
\btitle{Random effects models for combining results from controlled
and uncontrolled studies in a meta-analysis}.
\bjournal{J. Amer. Statist. Assoc.}
\bvolume{89}
\bpages{1523--1527}.
\bid{issn={0162-1459}, mr={1310241}}
\end{barticle}
%
\bptok{imsref}%
\endbibitem

\bibitem[\protect\citeauthoryear{Mila and Ngugi}{2011}]{Mila2011}
%
\begin{barticle}[author]
\bauthor{\bsnm{Mila}, \bfnm{A.~L.}\binits{A.~L.}} \AND
\bauthor{\bsnm{Ngugi}, \bfnm{H.~K.}\binits{H.~K.}}
(\byear{2011}).
\btitle{A Bayesian approach to meta-analysis of plant pathology studies}.
\bjournal{Phytopathology}
\bvolume{101}
\bpages{42--51}.
\bid{doi={10.1094/PHYTO-03-10-0070}, issn={0031-949X}, pmid={20822433}}
\end{barticle}
%
\bptok{imsref}%
\endbibitem

\bibitem[\protect\citeauthoryear{Owen}{2009}]{Owen2009}
%
\begin{barticle}[mr]
\bauthor{\bsnm{Owen},~\bfnm{Art~B.}\binits{A.~B.}}
(\byear{2009}).
\btitle{Karl {P}earson's meta-analysis revisited}.
\bjournal{Ann. Statist.}
\bvolume{37}
\bpages{3867--3892}.
\bid{doi={10.1214/09-AOS697}, issn={0090-5364}, mr={2572446}}
\end{barticle}
%
\bptok{imsref}%
\endbibitem

\bibitem[\protect\citeauthoryear{Pickrell et~al.}{2010}]{Pickrell2010}
%
\begin{barticle}[author]
\bauthor{\bsnm{Pickrell}, \bfnm{J.~K.}\binits{J.~K.}},
\bauthor{\bsnm{Marioni}, \bfnm{J.~C.}\binits{J.~C.}},
\bauthor{\bsnm{Pai}, \bfnm{A.~A.}\binits{A.~A.}},
\bauthor{\bsnm{Degner}, \bfnm{J.~F.}\binits{J.~F.}} \betal{et al.}
(\byear{2010}).
\btitle{Understanding mechanisms underlying human gene expression
variation with RNA sequencing}
\bjournal{Nature}
\bvolume{464}
\bpages{768--772}.
\end{barticle}
%
\bptok{imsref}%
\endbibitem

\bibitem[\protect\citeauthoryear{Servin and Stephens}{2008}]{servinstephens}
%
\begin{barticle}[author]
\bauthor{\bsnm{Servin}, \bfnm{B.}\binits{B.}} \AND
\bauthor{\bsnm{Stephens}, \bfnm{M.}\binits{M.}}
(\byear{2008}).
\btitle{Imputation-based analysis of association studies: Candidate
regions and quantitative traits}.
\bjournal{PLoS Genetics}
\bvolume{3}
\bpages{e114}.
\end{barticle}
%
\bptok{imsref}%
\endbibitem

\bibitem[\protect\citeauthoryear{Stangl and Berry}{2000}]{stangelberrybook}
%
\begin{bbook}[author]
\bauthor{\bsnm{Stangl}, \bfnm{D.~K.}\binits{D.~K.}} \AND
\bauthor{\bsnm{Berry}, \bfnm{D.~A.}\binits{D.~A.}}
(\byear{2000}).
\btitle{Meta-Analysis in Medicine and Health Policy}.
\bpublisher{Dekker}, \blocation{New York}.
\end{bbook}
%
\bptok{imsref}%
\endbibitem

\bibitem[\protect\citeauthoryear{Stephens}{2013}]{stephens2013}
%
\begin{barticle}[pbm]
\bauthor{\bsnm{Stephens},~\bfnm{Matthew}\binits{M.}}
(\byear{2013}).
\btitle{A unified framework for association analysis with multiple
related phenotypes}.
\bjournal{PLoS One}
\bvolume{8}
\bpages{e65245}.
\bid{doi={10.1371/journal.pone.0065245}, issn={1932-6203},
pii={PONE-D-13-00855}, pmcid={3702528}, pmid={23861737}}
\end{barticle}
%
\bptok{imsref}%
\endbibitem

\bibitem[\protect\citeauthoryear{Stranger et~al.}{2007}]{Stranger2007}
%
\begin{barticle}[author]
\bauthor{\bsnm{Stranger}, \bfnm{B.~E.}\binits{B.~E.}},
\bauthor{\bsnm{Nica}, \bfnm{A.~C.}\binits{A.~C.}},
\bauthor{\bsnm{Forrest}, \bfnm{M.~S.}\binits{M.~S.}},
\bauthor{\bsnm{Dimas}, \bfnm{A.}\binits{A.}},
\bauthor{\bsnm{Bird}, \bfnm{C.~P.}\binits{C.~P.}} \betal{et al.}
(\byear{2007}).
\btitle{Population genomics of human gene expression.}
\bjournal{Nat. Genet.}
\bvolume{39}
\bpages{1217--1224}.
\end{barticle}
%
\bptok{imsref}%
\endbibitem

\bibitem[\protect\citeauthoryear{Sutton and Abrams}{2001}]{Sutton2001}
%
\begin{barticle}[author]
\bauthor{\bsnm{Sutton}, \bfnm{A.~J.}\binits{A.~J.}} \AND
\bauthor{\bsnm{Abrams}, \bfnm{K.~R.}\binits{K.~R.}}
(\byear{2001}).
\btitle{Bayesian methods in meta-analysis and evidence synthesis}.
\bjournal{Stat. Methods Med. Res.}
\bvolume{10}
\bpages{277--303}.
\bid{issn={0962-2802}, pmid={11491414}}
\end{barticle}
%
\bptok{imsref}%
\endbibitem

\bibitem[\protect\citeauthoryear{Teslovich et~al.}{2010}]{Teslovich2010}
%
\begin{barticle}[author]
\bauthor{\bsnm{Teslovich}, \bfnm{T.~M.}\binits{T.~M.}},
\bauthor{\bsnm{Musunuru}, \bfnm{K.}\binits{K.}},
\bauthor{\bsnm{Smith}, \bfnm{A.~V.}\binits{A.~V.}},
\bauthor{\bsnm{Edmondson}, \bfnm{A.~C.}\binits{A.~C.}},
\bauthor{\bsnm{Stylianou}, \bfnm{I.~M.}\binits{I.~M.}} \betal{et al.}
(\byear{2010}).
\btitle{Biological, clinical and population relevance of 95 loci for
blood lipids}.
\bjournal{Nature}
\bvolume{466}
\bpages{707--713}.
\end{barticle}
%
\bptok{imsref}%
\endbibitem

\bibitem[\protect\citeauthoryear{Verzilli et~al.}{2008}]{Verzilli2008}
%
\begin{barticle}[author]
\bauthor{\bsnm{Verzilli}, \bfnm{C.~J.}\binits{C.~J.}},
\bauthor{\bsnm{Shah}, \bfnm{T.}\binits{T.}},
\bauthor{\bsnm{Casas}, \bfnm{J.~P.}\binits{J.~P.}},
\bauthor{\bsnm{Chapman}, \bfnm{J.}\binits{J.}},
\bauthor{\bsnm{Sandhu}, \bfnm{M.}\binits{M.}} \betal{et al.}
(\byear{2008}).
\btitle{Bayesian meta-analysis of genetic association studies with
different sets of markers}.
\bjournal{Am. J. Hum. Genet.}
\bvolume{82}
\bpages{859--872}.
\end{barticle}
%
\bptok{imsref}%
\endbibitem

\bibitem[\protect\citeauthoryear{Wakefield}{2009}]{Wakefield2009}
%
\begin{barticle}[mr]
\bauthor{\bsnm{Wakefield}, \bfnm{Jon}\binits{J.}}
(\byear{2009}).
\btitle{Bayes factors for genome-wide association studies: Comparison
with \mbox{$P$-}values}.
\bjournal{Genet. Epidemiol.}
\bvolume{33}
\bpages{79--86}.
\bid{doi={10.1002/gepi.20359}, issn={0741-0395}, pmid={18642345}}
\end{barticle}
%
\bptok{imsref}%
\endbibitem

\bibitem[\protect\citeauthoryear{Wen}{2011}]{wenthesis}
%
\begin{bmisc}[author]
\bauthor{\bsnm{Wen},~\bfnm{Xiaoquan}\binits{X.}}
(\byear{2011}).
\bhowpublished{Bayesian analysis of genetic association data,
accounting for heterogeneity.
Ph.D. thesis, Dept. Statistics, Univ. Chicago.}
\end{bmisc}
%
\bptok{imsref}%
\endbibitem

\bibitem[\protect\citeauthoryear{Wen and Stephens}{2014}]{suppA}
%
\begin{bmisc}[author]
\bauthor{\bsnm{Wen},~\bfnm{Xiaoquan}\binits{X.}} \AND
\bauthor{\bsnm{Stephens},~\bfnm{Matthew}\binits{M.}}
(\byear{2014}).
\bhowpublished{Supplement to ``Bayesian methods for genetic association analysis with
heterogeneous subgroups: From meta-analyses to gene--environment interactions.''
DOI:\doiurl{10.1214/13-AOAS695SUPP}}.
\bptok{imsref}%
\end{bmisc}
%
\endbibitem

\bibitem[\protect\citeauthoryear{Whitehead and
Whitehead}{1991}]{Whitehead1991}
%
\begin{barticle}[author]
\bauthor{\bsnm{Whitehead}, \bfnm{A.}\binits{A.}} \AND
\bauthor{\bsnm{Whitehead}, \bfnm{J.}\binits{J.}}
(\byear{1991}).
\btitle{A general parametric approach to the meta-analysis of
randomized clinical trials}.
\bjournal{Stat. Med.}
\bvolume{10}
\bpages{1665--1677}.
\bid{issn={0277-6715}, pmid={1792461}}
\end{barticle}
%
\bptok{imsref}%
\endbibitem

\bibitem[\protect\citeauthoryear{Willer, Li and Abecasis}{2010}]{Willer2010}
%
\begin{barticle}[author]
\bauthor{\bsnm{Willer}, \bfnm{Cristen J.}\binits{C.~J.}},
\bauthor{\bsnm{Li}, \bfnm{Yun}\binits{Y.}} \AND
\bauthor{\bsnm{Abecasis}, \bfnm{Gon{\c{c}}alo R.}\binits{G.~R.}}
(\byear{2010}).
\btitle{METAL: Fast and efficient meta-analysis of genomewide
association scans}.
\bjournal{Bioinformatics}
\bvolume{26}
\bpages{2190--2191}.
\bid{doi={10.1093/bioinformatics/btq340}, issn={1367-4811},
pii={btq340}, pmcid={2922887}, pmid={20616382}}
\end{barticle}
%
\bptok{imsref}%
\endbibitem

\end{thebibliography}
\end{document}